\documentclass[final,3p,times]{elsarticle}
   
\usepackage{amssymb}
\usepackage{xcolor,soul}
\usepackage{array, makecell} 
\usepackage{tabularx,booktabs}
\usepackage{graphicx}
\usepackage{multirow}
\usepackage{hhline}
\usepackage{amsmath}
\usepackage{caption}
\usepackage[labelformat=simple]{subcaption}
\usepackage{latexsym}
\usepackage{lineno}
\usepackage{hyperref}

\journal{npj Precision Oncology}

\begin{document}

\begin{frontmatter}


 
\title{An interpretable machine learning system for colorectal cancer diagnosis from pathology slides}

\author[inesc,feup,fn1,fn2]{Pedro C. Neto}
\author[imp,icbas,ipo1,fn1,fn3]{Diana Montezuma}
\author[inesc,feup,fn1,fn4]{Sara P. Oliveira}
\author[imp]{Domingos Oliveira}
\author[ipo2]{João Fraga}
\author[imp]{Ana Monteiro}
\author[imp]{João Monteiro}
\author[imp]{Liliana Ribeiro}
\author[imp]{Sofia Gonçalves}
\author[bern]{Stefan Reinhard}
\author[bern]{Inti Zlobec}
\author[imp]{Isabel M. Pinto}  
\author[inesc,feup]{Jaime S. Cardoso}

\fntext[fn1]{These authors contributed equally.}
\fntext[fn2]{pedro.d.carneiro@inesctec.pt}
\fntext[fn3]{diana.felizardo@impdiagnostics.com}
\fntext[fn4]{s.oliveira@nki.nl}
\affiliation[inesc]{organization={Institute for Systems and Computer Engineering, Technology and Science (INESC TEC)},
             addressline={R. Dr. Roberto Frias},
             city={Porto},
             postcode={4200-465},
             state={Porto},
             country={Portugal}}

 \affiliation[feup]{organization={Faculty of Engineering, University of Porto (FEUP)},
             addressline={R. Dr. Roberto Frias},
             city={Porto},
             postcode={4200-465},
             state={Porto},
             country={Portugal}}
 \affiliation[imp]{organization={IMP Diagnostics},
             addressline={Praça do Bom Sucesso, 61, sala 808},
            city={Porto},
            postcode={4150-146},
            state={Porto},
            country={Portugal}}
\affiliation[ipo1]{organization={Cancer Biology and Epigenetics Group, Research Center of IPO Porto (CI-IPOP) /
RISE@CI-IPOP (Health Research Network), Portuguese Oncology Institute of Porto (IPO Porto) / Porto Comprehensive Cancer Center (Porto.CCC)},
            addressline={R. Dr. António Bernardino de Almeida 865},
            city={Porto},
            postcode={4200-072},
            state={Porto},
            country={Portugal}}
\affiliation[ipo2]{organization={Department of Pathology, IPO-Porto},
            addressline={R. Dr. António Bernardino de Almeida 865},
            city={Porto},
            postcode={4200-072},
            state={Porto},
            country={Portugal}}
\affiliation[icbas]{organization={Doctoral Programme in Medical Sciences, School of Medicine and Biomedical
Sciences – University of Porto (ICBAS-UP)},
            addressline={R. Jorge de Viterbo Ferreira 228},
            city={Porto},
            postcode={4050-313},
            state={Porto},
            country={Portugal}}

\affiliation[bern]{organization={Institute of Pathology, University of Bern},
            addressline={Uni Bern, Murtenstrasse 31},
            city={Bern},
            postcode={3008},
            state={Bern},
            country={Switzerland}}




\begin{abstract}

Considering the profound transformation affecting pathology practice, we aimed to develop a scalable artificial intelligence (AI) system to diagnose colorectal cancer from whole-slide images (WSI). For this, we propose a deep learning (DL) system that learns from weak labels, a sampling strategy that reduces the number of training samples by a factor of six without compromising performance, an approach to leverage a small subset of fully annotated samples, and a prototype with explainable predictions, active learning features and parallelization. Noting some problems in the literature, this study is conducted with one of the largest WSI colorectal samples dataset with approximately 10,500 WSIs. Of these samples, 900 are testing samples. Furthermore, the robustness of the proposed method is assessed with two additional external datasets (TCGA and PAIP) and a dataset of samples collected directly from the proposed prototype. Our proposed method predicts, for the patch-based tiles, a class based on the severity of the dysplasia and uses that information to classify the whole slide. It is trained with an interpretable mixed-supervision scheme to leverage the domain knowledge introduced by pathologists through spatial annotations. The mixed-supervision scheme allowed for an intelligent sampling strategy effectively evaluated in several different scenarios without compromising the performance. On the internal dataset, the method shows an accuracy of 93.44\% and a sensitivity between positive (low-grade and high-grade dysplasia) and non-neoplastic samples of 0.997. On the external test samples varied with TCGA being the most challenging dataset with an overall accuracy of 84.91\% and a sensitivity of 0.996.

\end{abstract}



\begin{keyword}
Clinical Prototype \sep Colorectal Cancer \sep Interpretable Artificial Intelligence	\sep Deep Learning \sep Whole-Slide Images


\end{keyword}

\end{frontmatter}


\section*{Introduction}\label{sec_intro}

Colorectal cancer (CRC) incidence and mortality are increasing,  with projections indicating continued growth until at least 2040, according to estimations of the International Agency for Research on Cancer~\cite{IARC2021}. Nowadays, it is the third most incident (10.7\% of all cancer diagnoses) and the second most deadly type of cancer~\cite{IARC2021}. Despite the pessimist predictions, CRC is preventable and curable when detected in its earlier stages. Thus, effective screening through medical examination, imaging techniques and colonoscopy are of utmost importance~\cite{DICE2019, Hassan2020}.
Notwithstanding the CRC detection capabilities shown by imaging/endoscopic techniques, the definite diagnosis of cancer is always based on the pathologist's evaluation of the histological samples. The stratification of neoplasia development stages consists of non-neoplastic (NNeo), low-grade dysplasia (LGD), high-grade dysplasia (HGD, including intramucosal carcinomas), and invasive carcinomas, from the initial to the latest stage of cancer progression, respectively. In spite of the inherent subjectivity of the dysplasia grading system~\cite{Mahajan2013}, recent guidelines from the European Society of Gastrointestinal Endoscopy (ESGE), as well as those from the US multi-society task force on CRC, consistently recommend shorter surveillance intervals for patients with polyps with high-grade dysplasia, regardless of their dimension~\cite{Hassan2020, Gupta2020}. Hence, grading dysplasia is still routinely performed by pathologists worldwide when assessing colorectal tissue samples.

Private datasets of digitised slides are becoming widely available, in the form of whole-slide images (WSI), with an increase in the adoption of digital workflows~\cite{Eloy2021, Fraggetta2021,montezuma2022digital}. Whole-slide imaging eases the revision of old cases, data sharing and peer-review~\cite{Madabhushi2016,Rakha2020}. It has also created several research opportunities within the computer vision domain, especially due to the complexity of the problem and the high dimensions of WSIs~\cite{Veta2015, Campanella2019, Oliveira2020,Albuquerque2021}. Robust and high-performance systems can be valuable assets to the digital workflow of a laboratory, especially if they are transparent and interpretable~\cite{Madabhushi2016,Rakha2020}. However, some limitations still affect the applicability of such solutions into practice~\cite{Oliveira2021}. 

The analysis of CRC samples from WSI is divided into two different branches: classification of regions of interest, and classification of WSI. On the latter topic, despite the limitations, researchers have been improving the state of the art on the classification of the slide from individual tile classification, or aggregation methods~\cite{Thakur2020, Wang2020, Oliveira2021, davri2022deep}.
In 2020, Iizuka~\textit{et al.}~\cite{Iizuka2020} used a recurrent neural network (RNN) to aggregate the predictions of individual tiles processed by an Inception-v3 network into non-neoplastic, adenoma (AD) and adenocarcinoma (ADC). Due to the large dimensions of WSI related to their pyramidal format (with several magnification levels)~\cite{Tizhoosh2018}, usually over 50,000 $\times$ 50,000 pixels, it is usual to use a scheme consisting of a grid of tiles. This scheme permits the acceleration of the processing steps since the tiles are small enough to fit in the memory of the graphics processing units (GPU), popular units for the training of deep learning (DL). Wei~\textit{et al.}~\cite{Wei2020} studied the usage of an ensemble of five distinct ResNet networks, in order to distinguish the types of CRC adenomas H\&E stained slides. Song~\textit{et al.}~\cite{Song2020} experimented with a modified DeepLab-v2 network for tile classification, and proposed pixel probability thresholding to detect CRC adenomas. Both Xu~\textit{et al.}~\cite{Xu2020} and Wang~\textit{et al.}~\cite{Wang2021, Yu2021} looked into the performance of the Inception-v3 architecture to detect CRC, with the latter also retrieving a cluster-based slide classification and a map of predictions. The MuSTMIL~\cite{Marini2021} method \textcolor{black}{classifies} five colon-tissue findings: normal glands, hyperplastic polyps, low-grade dysplasias, high-grade dysplasias and carcinomas. This classification originates from a multitask architecture that leverages several levels of magnification of a slide. Ho~\textit{et al.}~\cite{Ho2022} extended the experiments with multitask learning, but instead of leveraging the magnification, its model aims to jointly segment glands, detect tumour areas and sort the slides into low-risk (benign, inflammation or reactive changes) and high-risk (adenocarcinoma or dysplasia) categories. The architecture of this model is considerably more complex, with regard to the number of parameters, and is known as Faster-RCNN with a ResNet-101 backbone network for the segmentation task. Further to this task, a gradient-boosted decision tree completes the pipeline that results in the final grade. More recently, Bokhorst~\textit{et al.}~\cite{bokhorst2023} presented an DL-based method to segment multiple colorectal tissue compartments and then used the best performing model classify biopsies as either (1) high-risk (tumor and high-grade dysplasia), (2) low-grade dysplasia, (3) hyperplasia and (4) benign; achieving an one-vs-all AUC of 0.87 for the high-risk category. Notably, Graham~\textit{et al.}~\cite{Graham2023} have developed a graph neural network, Interpretable Gland-Graphs using a Neural Aggregator (IGUANA), to distinguish colorectal samples in normal vs. abnormal (non-neoplastic and neoplastic), achieving a sensitivity threshold of 99\%, proposing, with their model, to reduce the number of normal slides to be reviewed by pathologists by 55\%. 

Our work aims to further contribute to the landscape of computer-aided diagnosis (CAD) systems for colorectal pathology, addressing current hurdles and limitations:
- The high volume of data needed, in addition to the massive resolution of the images, creates a significant bottleneck of DL approaches that extract patches from the whole slides. Hence, we introduce an efficient sampling approach that is performed once without sacrificing predictive performance on the classification. Leveraging the domain knowledge introduced in the data, by the expert pathologists, in the form of annotations at the pixel level, the model is capable of predicting pseudo-labels for the non-annotated samples. Leveraging these new pseudo-labels, we can discard tiles with the least meaningful pseudo-labels, resulting in 6$\times$ less tiles while retaining most of the important information. This process is preceded by a supervised learning step using the pixel level annotations where the model learns how to create the pseudo-labels for the sampling step. After, the sampling is followed by a weakly-supervised approach on the reduced set of slides and using only slide labels.
Our dataset contains, approximately, 10500 high-quality slides from IMP Diagnostics. A large part of this dataset is publicly available~\cite{neto2024dataset}\textcolor{black}{, with corresponding case diagnostic labels} (making it one of the largest colorectal samples (CRS) datasets available to date).  We validate our proposed model in two different external datasets that vary in quality, country of origin and laboratory, ensuring its generalisation capability and robustness. 
Importantly, in order to bring this CAD system into production, and to infer its usefulness within clinical practice, we developed a prototype, with explainable predictions (visual maps), that was tested and evaluated by pathologists.

To summarise, in this paper we propose a novel dataset with more than thirteen million tiles, a sampling approach to reduce the difficulty of using large datasets, an accurate DL model that is trained with mixed supervision, is evaluated on four datasets, and finally incorporated in a prototype that provides a simple integration in clinical practice and visual explanations of the model's predictions. This way, we are a step closer to making CAD tools a reality for colorectal diagnosis.

\section*{Results}
\label{sec:results}

In this section, the results are organised to first demonstrate the effectiveness of sampling, followed by an evaluation of the model in the two internal datasets (CRS10K and the prototype dataset), and in the external datasets.

\subsection*{On the effectiveness of sampling}

\begin{figure}[h!]
    \centering
    \includegraphics[width=0.8\textwidth]{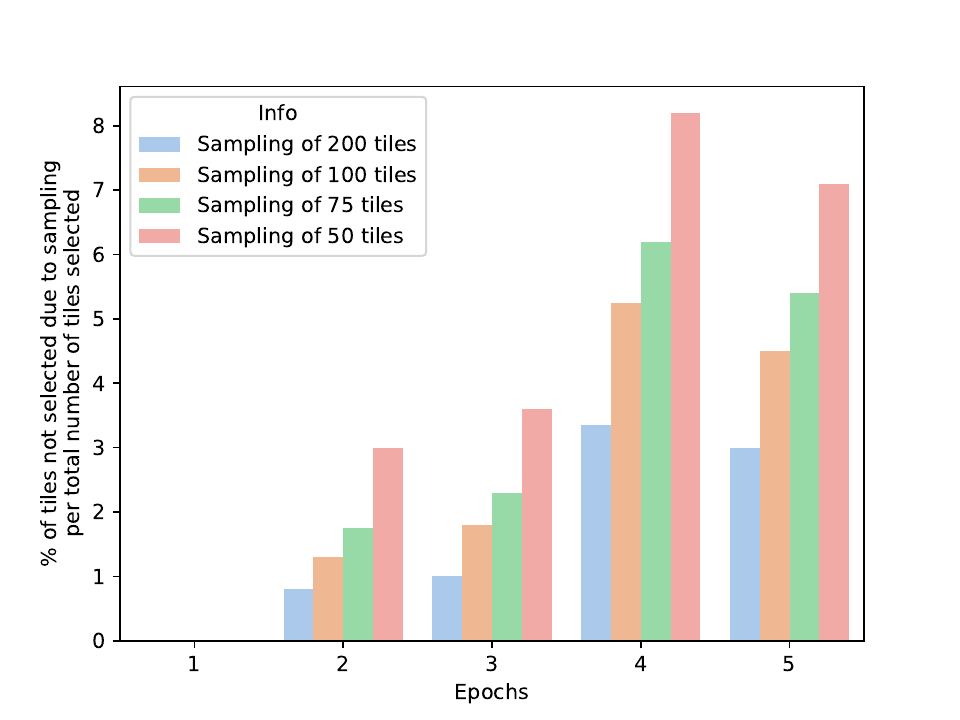}
    \caption{\textbf{Tile sampling impact on information loss}: percentage of tiles not selected due to sampling with different thresholds, over the first four inference epochs. The blue bar represents a sampling strategy that retains 200 tiles per slide, the orange bar is for a strategy that retains 100 tiles, the green bar represents a strategy that retains 75 tiles and finally the strategy represented by the red line retains 50 tiles per slide.  }
    \label{fig:sampling_results}
\end{figure} 

To find the most suitable threshold for sampling the tiles used in the weakly supervised training, we evaluated the percentage of relevant tiles that would be left out of the selection, if the original set was reduced to 75, 100, 150 or 200 tiles, over the first five inference epochs. A tile is considered relevant if it shares the same label as the slide, or if it would take part in the learning process in the weakly-supervised stage. As it is possible to see in Figure~\ref{fig:sampling_results}, if we set the maximum number of tiles to 200 after the second loop of inference, we would discard only 3.5\% of the potentially informative tiles, in the worst-case scenario. On the other side of the spectrum, a more radical sampling of only 50 tiles would lead to a cut of up to 8\%.

Moreover, to assess the impact of this sampling on the model’s performance, we also evaluated the accuracy and the QWK with and without sampling the top 200 tiles after the first inference iteration (Table~\ref{tab:sampling_comparision_results}). This evaluation considered sampling applied only to the training tile set, and to both the training and validation tile sets. As can be noticed, the performance is not degraded and the model is trained in a much faster way. In fact, using the setup previously mentioned, the first epoch of inference, with the full set of tiles takes 28h to be completed, while from the second loop the training time decreases to only 5h per epoch. Without sampling, training the model for 50 epochs would take around 50 days, whereas with sampling it takes around 10.

\begin{table}[htbp]
\footnotesize	
\centering
\caption{Model performance comparison with and without tile sampling of the top 200 tiles from the first inference iteration. Compared the best epoch of the initial five epochs and of the initial ten epochs. Validation is represented as Val and the best results are in bold.}
\label{tab:sampling_comparision_results}
\begin{tabular}{lcccc}
\toprule
\cmidrule{2-5} Sampling & Best ACC at 5th epoch & Best ACC at 10th epoch &  Best QWK at 5th epoch & Best QWK at 10th epoch  \\
\midrule
No & $84.94\% \pm 2.20$ & $86.42\% \pm 2.11$ & $0.809 \pm 0.024$ & $\textbf{0.829} \pm \textbf{0.023}$ \\
Train & $85.43\% \pm 2.18$ & $86.82\% \pm 2.08$ & $0.817 \pm 0.024$ & $0.828 \pm 0.023$ \\
Train and Val. & $\textbf{86.12\%} \pm \textbf{2.13}$ & $\textbf{86.92\%} \pm \textbf{2.08}$  & $\textbf{0.824} \pm \textbf{0.023}$ &$\textbf{0.829} \pm \textbf{0.023}$  \\
\bottomrule
\end{tabular}
\end{table}

\subsection*{CRS10K and Prototype} \label{subsec:CRC}

CRS10K test set and the prototype dataset were collected through different procedures. The first followed the same data collection process as the complete dataset, whereas the second originated from routine samples. Thus, the evaluation of both these sets is done separately.

\begin{table}[h!]
\centering
\footnotesize	
\caption{Model performance evaluation on the CRS10K test set. The binary accuracy is calculated as NNeo vs all. Accuracy is represented as (ACC). In bold are the best results per column.}
\label{tab:results_test_set}
\begin{tabular}{lccc}
\toprule
Method  & ACC & Binary ACC & Sensitivity \\ \midrule
iMIL4Path &  $91.33\%\pm1.84$  & $97.00\%\pm1.11$  & $\textbf{0.997}\pm\textbf{0.004}$\\
Ours (CRS4K) &  $89.44\%\pm2.01$  &$96.11\%\pm1.26$  & $\textbf{0.997}\pm \textbf{0.004}$ \\
Ours (CRS10K) wo/ Agg & $\textbf{93.44\%}\pm\textbf{1.62}$ & $\textbf{97.78\%}\pm \textbf{0.96}$  & $0.996\pm 0.005$\\
Ours (CRS10K) w/ Agg & $90.67\% \pm 1.90 $ & $97.55\% \pm 1.01$& $0.985\pm 0.009$ \\
\bottomrule
\end{tabular}
\end{table}

The first experiment was conducted on the CRS10K test set. As expected, the steep increase in the number of training samples led to a significantly better algorithm in this test set. Initially, the model trained on the CRS10K correctly predicted the class of 819 out of 900 samples. For the wrong 81 cases, the pathologists performed a blind review and found that the algorithm was indeed correct in 22 of them. This led to a correction in the labels of the test set, and the appropriate adjustment of the metrics.  In Table~\ref{tab:results_test_set}, the performance of the different algorithms is presented. CRS10K outperforms the other approaches by a reasonable margin. 

\begin{table}[h!]
\centering
\footnotesize	
\caption{$\mathcal{X}^2$ and p-value computed using the McNemar's test for the models evaluated on the Test set. If  $\mathcal{X}^2$  is $>$ than 3.84 the difference between two methods is statistically significant. These statistically significant differences are highlighted in bold. P-value, under parentheses,  is computed by calculating the area under the PDF of the chi squared distribution to the right of  $\mathcal{X}^2$.}
\label{tab:test_stats_eval}
\begin{tabular}{lccc}
Method  & iMIL4Path &Ours (CRS4K) & Ours (CRS10K) wo/ Agg \\ \midrule
iMIL4Path &  -  & 1.82 (0.177) & 1.92 (0.166) \\
Ours (CRS4K) & 1.82 (0.177)&- &  \textbf{6.94 (0.008)} \\
Ours (CRS10K) wo/ Agg & 1.92 (0.166) & \textbf{6.94 (0.008)} &- \\
\bottomrule
\end{tabular}
\end{table}

\textcolor{black}{Using the McNemar's test, it was shown that there were significant different performances between the proposed model trained on CRS10K data and the model trained on CRS4K with a p-value of 0.008(Table~\ref{tab:test_stats_eval}). The  differences between the proposed methods trained on CSR10K and CSR4K, and iMIL4Path are not statistically significant with p-value of 0.166 and 0.177 respectively.} We further applied the aggregation proposed by Neto~\textit{et al.}~\cite{neto2022imil4path} to the proposed method trained on CRS10K, but without gains in performance. Despite being trained on the same dataset, iMIL4Path and the proposed methodology trained on CRS4K, they utilise different splits for training and validation, as well as different optimisation techniques due to the deterministic approach.

\begin{figure}[h!]
    \centering
    \includegraphics[width=.9\textwidth]{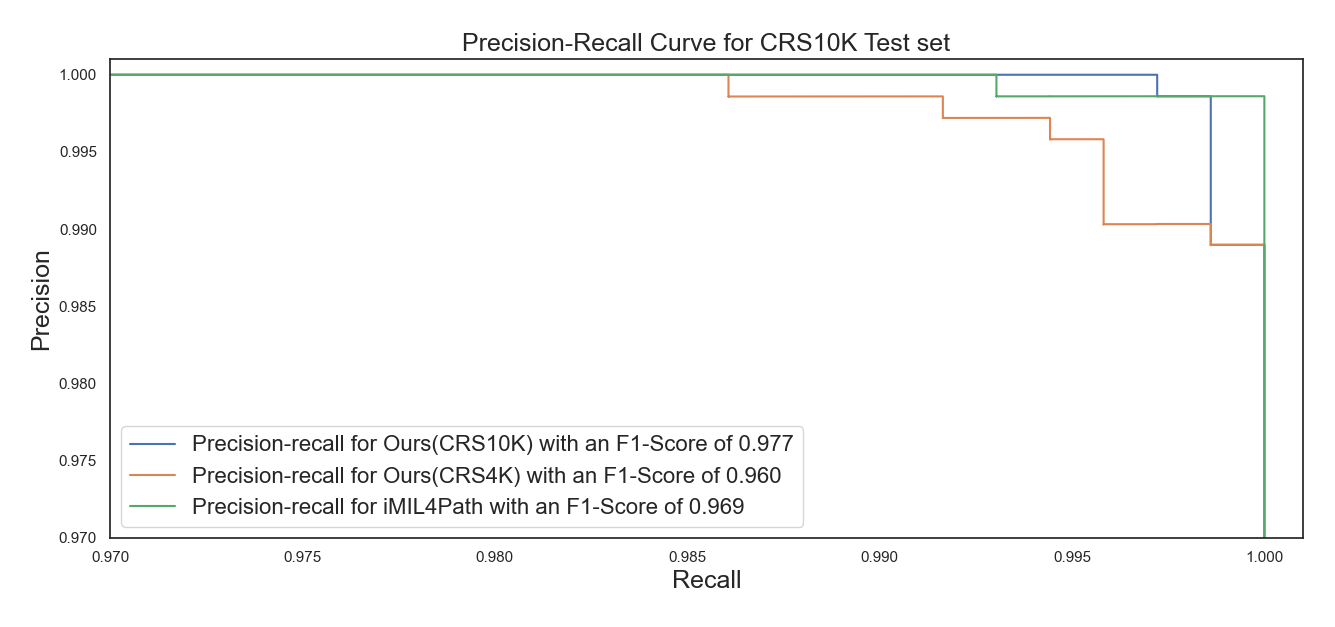}
    \caption{\textbf{Precision-recall curve on the on the CRS10K test set:} For the three distinct models, we have calculated the Precision-recall curve on this dataset. Includes an indication of the F1-Score for each of the different models. The blue line represents the curve of Our method when trained on CRS10K, while the orange line shows the same method when trained on CRS4K. The green line is the curve of iMIL4Path.}
    \label{fig:10k_prc}
\end{figure}

A more in-depth inspection of the performance considering the different errors is shown in Figure~\ref{fig:10k_prc}, where the precision-recall curves for the three models is shown. Moreover, the F1-Score is also included, which shows that the most balanced model is the one that we proposed.

\begin{figure}[h!]
    \centering
    \includegraphics[width=\textwidth]{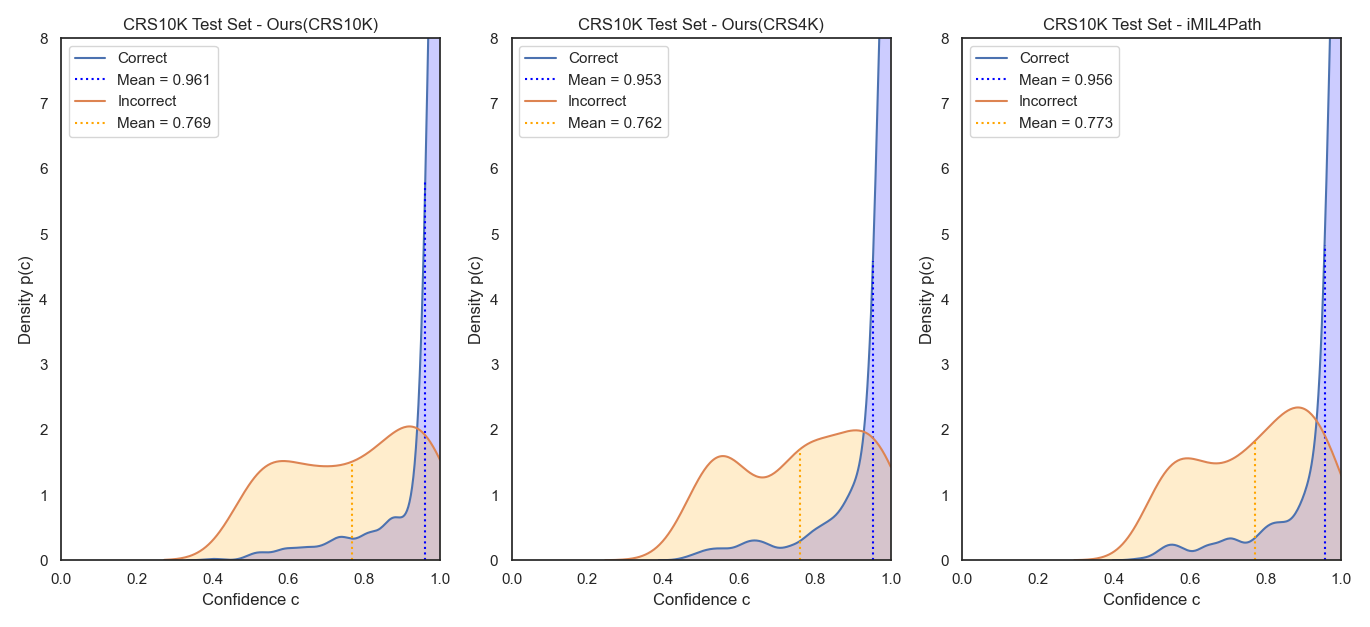}
    \caption{\textbf{Confidence analysis for correct and incorrect predictions on the CRS10K test set:} Kernel density estimation of the confidences of correct and incorrect predictions performed on the three-class classification problem by three distinct models on the CRS10K test set. The plots represent, from left to right, the proposed method trained on CRS10K, the proposed method trained on CRS4K and iMIL4Path. In each plot, the blue line defines the density function of the correct samples and the blue dashed line the mean confidence of those samples. On the other hand, the orange solid and dashed lines represent the same for incorrect predictions.}
    \label{fig:test_set_confidences}
\end{figure} 

In addition to examining quantitative metrics, such as the accuracy of the model, we extended our study to include an analysis of the confidence in the model when it correctly predicts a class and when it makes an incorrect prediction. To this end, we recorded the confidence of the model for the predicted class and divided it into the set of correct and incorrect predictions. These were then used to fit a kernel density estimator (KDE). Figure~\ref{fig:test_set_confidences} shows the density estimation of the confidence values for the three different models. It is worth noting that, when correct, the model trained on the CRS10K, returns higher confidence levels as shown by the shift of its mean towards values close to one. On the other hand, the confidence values of its incorrect predictions decrease significantly, and although it does not present the lowest values, it shows the largest gap between correct and incorrect means.

\begin{table}[h!]
\centering
\footnotesize	
\caption{Model performance evaluation on the prototype test set. Accuracy is represented as (ACC). The binary accuracy is calculated as NNeo vs all. In bold are the best results per column.}
\label{tab:prototype_comparison_results}
\begin{tabular}{lccc}
\toprule
Method  & ACC & Binary ACC & Sensitivity \\ \midrule
iMIL4Path &  $\textbf{89.00}\%\pm \textbf{6.13}$  & $96.00\%\pm 3.84$  & $\textbf{1.000}\pm \textbf{0.000}$\\
Ours (CRS4K) &  $85.00\%\pm 6.99$  &$93.00\%\pm 5.00$  & $\textbf{1.000}\pm \textbf{0.000}$ \\
Ours (CRS10K) wo/ Agg & $\textbf{89.00}\%\pm \textbf{6.13}$ & \textbf{$\textbf{98.00}\%\pm \textbf{2.74}$}  & \textbf{$0.986\pm 0.026$}\\
Ours (CRS10K) w/ Agg &$85.00\% \pm 6.99 $ & \textbf{$\textbf{98.00}\%\pm \textbf{2.74}$} & \textbf{$0.986\pm 0.026$}\\
\bottomrule
\end{tabular}
\end{table}

When tested on the prototype data (n=100), the importance of a higher volume of data remains visible (Table~\ref{tab:prototype_comparison_results}). Nonetheless, the performance of iMIL4Path~\cite{neto2022imil4path} approach is comparable to the proposed approach trained on CRS10K. It is worth noting that the latter achieves better performance on the binary accuracy at the cost of a decrease in sensitivity. In other words, the capability to detect negatives increases significantly.

\begin{table}[h!]
\centering
\footnotesize	
\caption{$\mathcal{X}^2$ and p-value computed using the McNemar's test for the models evaluated on the Prototype set. If  $\mathcal{X}^2$  is $>$ than 3.84 the difference between two methods is statistically significant. These statistically significant differences are highlighted in bold. P-value, under parentheses,  is computed by calculating the area under the PDF of the chi squared distribution to the right of  $\mathcal{X}^2$.}
\label{tab:proto_stats_eval}
\begin{tabular}{lccc}
Method  & iMIL4Path &Ours (CRS4K) & Ours (CRS10K) wo/ Agg \\ \midrule
iMIL4Path &  -  & 0.13 (0.718) & 0.00 (1.000) \\
Ours (CRS4K) &  0.13 (0.718) &- & 0.30 (0.584) \\
Ours (CRS10K) wo/ Agg & 0.00 (1.000)& 0.30 (0.584) &- \\
\bottomrule
\end{tabular}
\end{table}

\textcolor{black}{The McNemar's test did show significant differences between any of the methods  (Table~\ref{tab:proto_stats_eval}).} 
Similar performance drops were linked with the introduction of aggregation.

\begin{figure}[h!]
    \centering
    \includegraphics[width=\textwidth]{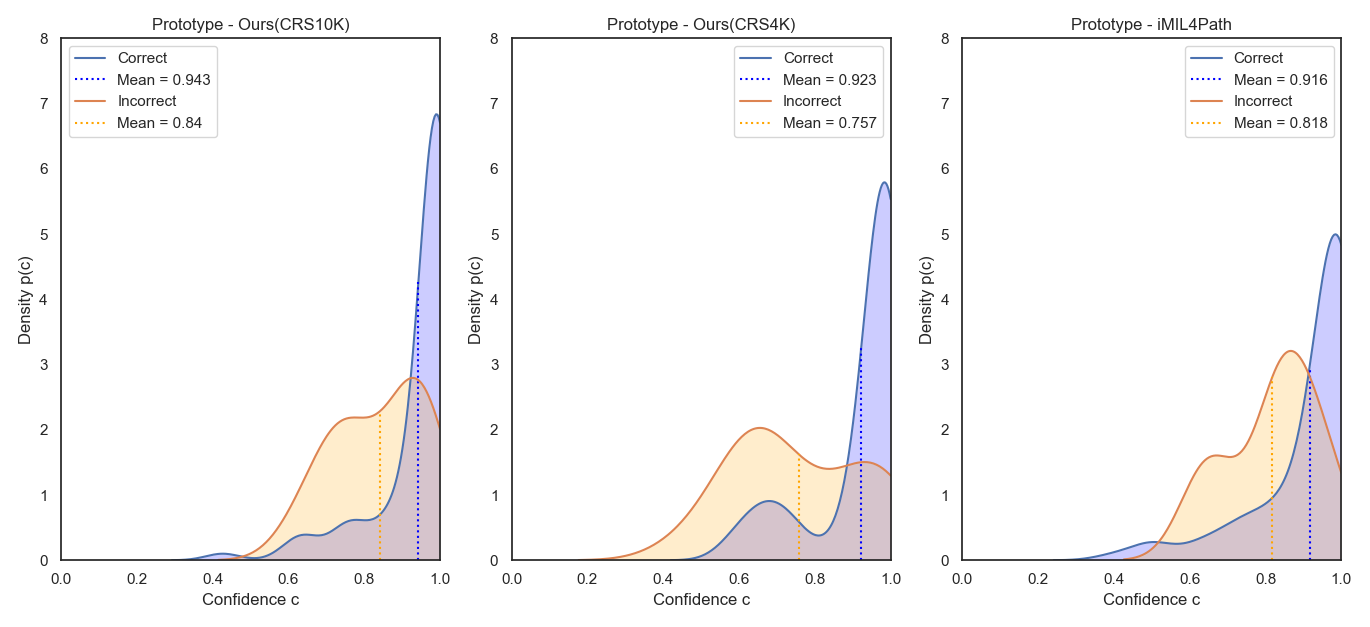}
    \caption{\textbf{Confidence analysis for correct and incorrect predictions on the Prototype set:} Kernel density estimation of the confidences of correct and incorrect predictions performed on the three-class classification problem by three distinct models on the prototype set. The plots represent, from left to right, the proposed method trained on CRS10K, the proposed method trained on CRS4K and iMIL4Path. In each plot, the blue line defines the density function of the correct samples and the blue dashed line the mean confidence of those samples. On the other hand, the orange solid and dashed lines represent the same for incorrect predictions.}
    \label{fig:prototype_confidences}
\end{figure}

Despite similar results, the confidence of the model in its predictions is distinct in all three approaches, as seen in Figure~\ref{fig:prototype_confidences}. The proposed approach when trained on the CRS10K dataset has a larger density on values close to one when the predictions are correct, and the mean confidence of those predictions is, once more, higher than the other approaches. However, especially when compared to the proposed approach trained on the CRS4K, the confidence of wrong predictions is also higher. It can be a result of a larger set of wrong predictions available on the latter approach. Nonetheless, the steep increase in the density of values closer to one further indicates that there is room to explore other effects of extending the number of training samples, besides benefits in quantitative metrics.

\subsection*{Domain Generalisation Evaluation} \label{subsec:TCGA}

To ensure the generalisation of the proposed approach across external datasets, we have evaluated their performance on TCGA and PAIP datasets. Moreover, we conducted a similar analysis to both of these datasets, as the one done for the internal datasets.

\begin{table}[h!]
\centering
\footnotesize	
\caption{Model performance evaluation on the PAIP test set. The binary accuracy is calculated as NNeo vs all. Accuracy is represented as (ACC). In bold are the best results per column.}
\label{tab:paip_comparison_results}
\begin{tabular}{lccc}
\toprule
Method  & ACC & Binary ACC & Sensitivity \\ \midrule
iMIL4Path &  $99.00\% \pm 1.95$  & $\textbf{100.00\%}\pm \textbf{0.00}$  & $\textbf{1.000}\pm \textbf{0.000}$\\
Ours (CRS4K) &  $69.00\%\pm 9.06$  &$\textbf{100.00\%}\pm \textbf{0.00}$  & $\textbf{1.000}\pm \textbf{0.000}$ \\
Ours (CRS10K) wo/ Agg & $\textbf{100.00\%}\pm \textbf{0.00}$ & $\textbf{100.00\%}\pm \textbf{0.00}$  & $\textbf{1.000}\pm \textbf{0.000}$\\
Ours (CRS10K) w/ Agg & $52.00 \pm 9.79$ & $\textbf{100.00\%}\pm \textbf{0.00}$  & $\textbf{1.000}\pm \textbf{0.000}$\\
\bottomrule
\end{tabular}
\end{table}

From the two datasets, PAIP is arguably the closest to CRS10K. It contains similar tissue, despite its colour differences. The performances of the proposed approaches were expected to match the performance of iMIL4Path in this dataset. However, it did not happen for the version trained on the CRS4K dataset, as seen in Table~\ref{tab:paip_comparison_results}. A possible explanation concerns potential overfitting to the training data potentiated by an increase in the number of epochs of fully and weakly supervised training, a slight decrease in the tile variability in the latter approach, and a smaller number of samples when compared to the version trained on CRS10K. This version, trained on the larger set, mitigates the problems of the other method due to a significant increase in the training samples.  Moreover, it is worth noting that in all three approaches, the errors corresponded only to a divergence between low and high-grade cases, with no non-neoplastic cases being classified as high-grade or vice-versa. As in previous sets, the version trained on the CRS10K dataset outperforms the remaining approaches. Using aggregation in this dataset leads to a discriminative power to distinguish between high- and low-grade lesions that is close to random.

\begin{table}[h!]
\centering
\footnotesize	
\caption{$\mathcal{X}^2$ and p-value computed using the McNemar's test for the models evaluated on the TCGA set. If  $\mathcal{X}^2$  is $>$ than 3.84 the difference between two methods is statistically significant. These statistically significant differences are highlighted in bold. P-value, under parentheses,  is computed by calculating the area under the PDF of the chi squared distribution to the right of  $\mathcal{X}^2$.}
\label{tab:tcga_stats_eval}
\begin{tabular}{lccc}
Method  & iMIL4Path &Ours (CRS4K) & Ours (CRS10K) wo/ Agg \\ \midrule
iMIL4Path &  -  & 0.04 (0.839)  & \textbf{26.26 (0.000)}  \\
Ours (CRS4K) &  0.04  (0.839) &- & \textbf{31.03 (0.000)}  \\
Ours (CRS10K) wo/ Agg & \textbf{26.26 (0.000)} & \textbf{31.03 (0.000)} &- \\
\bottomrule
\end{tabular}
\end{table}

\textcolor{black}{The McNemar's test indicated a significant difference in the performance difference between the model trained on CRS10K and the one trained on the CRS4k (p-value of 0.00), and between the latter and iMIL4Path (p-value of 0.00). However, there was no significant difference between iMIL4Path and the former with a p-value of 1.00 (~\ref{tab:tcga_stats_eval})} The confidence of the model was also calculated for this dataset~(Supplementary Figure 2), showing a visible shift towards higher values of confidence in the proposed approach trained on the CRS10K when compared to the method of iMIL4Path. The version trained on CRS4K showed very little separability between the confidence of correct and incorrect predictions.

\begin{table}[h!]
\centering
\footnotesize	
\caption{Model performance evaluation on the TCGA test set. The binary accuracy is calculated as NNeo vs all. Accuracy is represented as (ACC). In bold are the best results per column.}
\label{tab:tcga_comparison_results}
\begin{tabular}{lccc}
\toprule
Method  & ACC & Binary ACC & Sensitivity \\ \midrule
iMIL4Path&  $71.55\%\pm 5.80$  & $80.60\%\pm 5.05$  & $0.805\pm 0.051$\\
Ours (CRS4K) wo/ Agg & $70.69\%\pm 5.86$  & $98.71\%\pm 1.45$ & $0.991\pm 0.012$\\
Ours (CRS10K) wo/ Agg & $\textbf{84.91}\%\pm \textbf{4.61}$ & $\textbf{99.13\%}\pm \textbf{1.19}$ & $\textbf{0.996}\pm \textbf{0.008}$ \\
Ours (CRS10K) w/ Agg & $69.83\% \pm 5.91$ & $97.41\% \pm 2.04$ & $0.983 \pm 0.017$  \\
\bottomrule
\end{tabular}
\end{table}

The TCGA dataset has established itself as the most challenging for the proposed approaches. Besides the expected differences in colour and other elements, this dataset is mostly composed of resection samples, which are not present in the training dataset. As such, this presents itself as an excellent dataset to assess the capability of the model to handle these different types of samples. Both iMIL4Path and the proposed method trained on CRS4K have shown substantial problems in correctly classifying TCGA slides, as shown in Table~\ref{tab:tcga_comparison_results}. \textcolor{black} {This can be explained as in the TCGA dataset the majority of the high-grade lesions exhibit an invasive component, and the morphology of the tumoral lesions is altered with the invasiveness. Also, features like an abundance of desmoplastic stroma tend to manifest more prominently in the deeper regions of the tumor, as opposed to the superficial sections typically obtained through biopsy/polypectomy. These aspects also hold relevance in explaining the comparatively inferior outcomes observed in the TCGA dataset.} Despite having a lower performance on the general accuracy, the binary accuracy shows that our proposed method trained on CRS4K has much lower misclassification errors regarding the classification of high-grade samples as normal, demonstrating higher robustness of the new training approach against errors with a gap of two classes. As with other datasets, the proposed approach trained on CRS10K shows better results, this time by a significant margin with no overlapping between the confidence intervals. 

\begin{table}[h!]
\centering
\footnotesize	
\caption{ $\mathcal{X}^2$ and p-value computed using the McNemar's test for the models evaluated on the PAIP set. If  $\mathcal{X}^2$  is $>$ than 3.84 the difference between two methods is statistically significant. These statistically significant differences are highlighted in bold. P-value, under parentheses,  is computed by calculating the area under the PDF of the chi squared distribution to the right of  $\mathcal{X}^2$.}
\label{tab:paip_stats_eval}
\begin{tabular}{lccc}
Method  & iMIL4Path &Ours (CRS4K) & Ours (CRS10K) wo/ Agg \\ \midrule
iMIL4Path &  -  & \textbf{26.28 (0.000)}  & 0.00 (1.000) \\
Ours (CRS4K) &  \textbf{26.28 (0.000)}  &- &  \textbf{29.03 (0.000)} \\
Ours (CRS10K) wo/ Agg & 0.00 (1.000) & \textbf{29.03 (0.000)}  &- \\
\bottomrule
\end{tabular}
\end{table}

\textcolor{black}{This was further confirmed by the McNemar's test which once more highlighting the better performance of the proposed model with p-values of 0.00 when compared to either iMIL4Path or the same model trained on CRS4K. The lack of significance between the differences of iMIL4Path and the model trained on CRS4K (p-value of 0.839) further emphasises the capability of the sampling strategy to retain the results (~\ref{tab:paip_stats_eval}).} The confidence predictions for the three models were also assessed (~(Supplementary Figure 3), indicating a behaviour in line with the accuracy-based performance. Also, the model trained on CRS10K showed a shift of wrong predictions' confidence towards smaller values, indicating that it is possible to quantify the uncertainty of the model and avoid the majority of the wrong predictions. In other words, when the uncertainty is above a learnt threshold, then the model refuses to make any prediction which is extremely useful in models designed as a second opinion system.

\subsection*{Reject option}

Following the confidence analysis previously introduced, we further explore the possibility of rejecting some samples that represent lower levels of confidence.

\begin{figure}[h!]
    \centering
    \includegraphics[width=\textwidth]{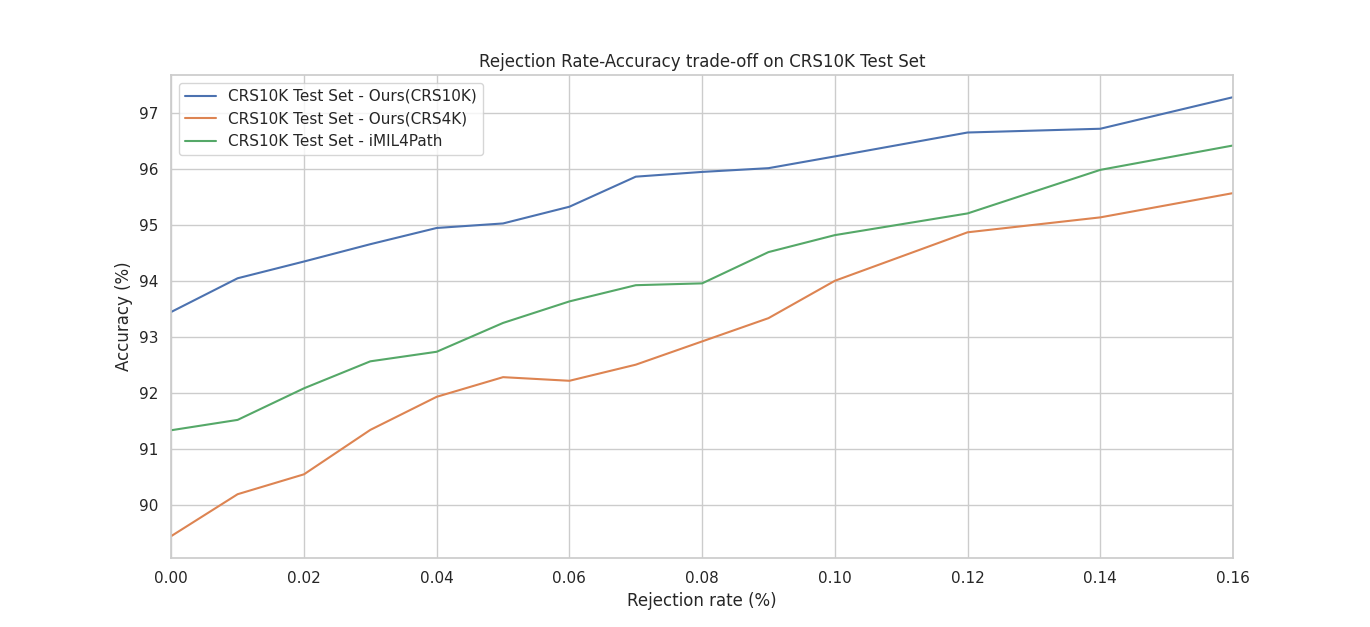}
    \caption{\textbf{Accuracy-vs-Rejection-rate for the models evaluated on the CRS10K test set}. Relation between the accuracy and the percentage of samples not classified by the model. Both axes are in percentage. The blue line represents Our method when trained on CRS10K, while the orange line shows the same method when trained on CRS4K. The green line is for iMIL4Path}
    \label{fig:10k_rejection}
\end{figure}

On the CRS10K test set, the reject rate correlates with an improved performance on all the algorithms displayed on Figure~\ref{fig:10k_rejection}. On our proposed algorithm we achieve 1.5 percent points improvement at a rejection rate of 4\% resulting in a accuracy of approximately 95\%. Moreover, if we reject 16\% of the samples (i.e. still reducing the pathologist workload by 84\%) the accuracy of the model is of 97.27\%. With a rejection rate of 50\%, which is less beneficial to pathologists, the accuracy would rise to 99.48\%.
The possibility of a reject option was also explored for the prototype dataset and TCGA dataset~(Supplementary Figures 4 and 5). We have not conducted this study on the PAIP dataset because the performance was already around 100\% in two of the main algorithms evaluated.

\subsection*{Prototype usability in clinical practice}

As it is currently designed, the algorithm works preferentially as a ``second opinion", allowing the assessment of difficult and troublesome cases, without the immediate need for the intervention of a second pathologist. Due to its "user-friendly" interface, the cases can be easily introduced into the system and results are rapidly shown and accessed. Also, by presenting visualisation maps, the pathologist is able to compare his own remarks to those of the algorithm itself, towards a future "AI-assisted diagnosis", where the pathologist has a pivotal role. Further, the prototype allows for user feedback (agreeing or not with the model's proposed result), which can be integrated into further updates of the software and could be leveraged in the future to feature active learning. Also interesting, \textcolor{black}{would be} the possibility of using \textcolor{black}{such a} prototype as a triage system on a pathologist's daily workflow by running upfront, before the pathologist checks the cases. Signalling the cases that would need to be more urgently observed (namely high-risk lesions) would allow the pathologists to prioritise their workflow. Further, by providing a previous assessment of the cases, it could also contribute to enhancing the pathologists' efficiency. \textcolor{black}{As such, this is one of our future work objectives.} Presently, there is no recommendation for dual independent diagnosis of colorectal biopsies (contrary to gastric biopsies, where, in cases that surgical treatment is considered, it is recommended to obtain a pre-treatment diagnostic second opinion ~\cite{WHO2019}), but, in the future, this can also become a requirement for colorectal samples. As such, CAD systems to assist pathologists in colorectal diagnosis can become even more important, being their relevance further amplified due to the worldwide shortage of pathologists.

\section*{Discussion}
\label{sec:conclusion}

In this work, we have proposed a redesig of the previous MIL methodology applied to colorectal cancer diagnosis. We aimed to develop a scalable, efficient and interpretable solution for this task. For this, we have worked on a mixed supervision approach to design a sampling strategy, which utilises the knowledge from the full supervision training as a proxy to tile utility. Secondly, we studied the confidence that the model shows in its predictions. Our target in this latter part was to infer the possibility of using a reject option based on the confidence of the model. 
The results have shown that this confidence has the potential to be a resource to quantify uncertainty and avoid wrong predictions on low-certainty scenarios. The model was entirely integrated within a web-based prototype to assist pathologists in their routine work.

The proposed methodology was evaluated on several datasets, including two external sets. Through this evaluation, it was possible to infer that the performance of the proposed methodology benefits from a larger dataset and surpasses the performance of previous state-of-the-art models that were evaluated on this benchmark. As such, and given the excelling results that originated from the increase in the dataset, we are also publicly releasing the majority of the CRS10K dataset \textcolor{black}{WSIs and case diagnostic labels}, one of the largest publicly available colorectal datasets composed of H\&E images in the literature, including the test set for the benchmark of distinct approaches across the literature~\cite{neto2024dataset}. 

Our findings have several noteworthy elements. First, we have shown that despite the ability to lead to better models, increasing the dataset size can be a double-edged sword due to the computational requirements of MIL solutions. With this in mind, and while conducting this study on one of the largest datasets for CRS, we have devised a sampling strategy that seems to minimize the information lost during training, leading to a comparable performance at 6x less processing time. Our method has also demonstrated the power of having a small portion of the dataset annotated to initialize the weights of the MIL model. We have further shown, that models trained on larger datasets seem to approximate more stable confidence distributions, leading to the possibility of using a reject option to comply with clinical requirements on the performance of the model. Finally, we have highlighted an interpretability method that is integrated into our prototype and supports the decision process of pathologists. 

Within the evaluation of datasets collected on similar configurations to the training data, the performance of the proposed model represents a step towards better algorithms for colorectal pathology. The high sensitivity did not compromise the overall accuracy. On datasets that originate from other centres and scanners, the performance was around 100\% of accuracy in one and around 85\% on the other, with the latter coming from completely different tissue samples. The comparison with other studies is highly limited by the test data. In our scenario, we have tested in a pool of  1332 samples, which is larger than several studies' train sets.  As we are releasing our test dataset, further research methods can be easily compared through it. 

We can note that the strong performance of the model, aligned with the prototype and the prediction maps, supports the utilisation of such a system as a second opinion within the routine process in a pathology lab, assisting pathologists in their daily routine, ensuring higher quality and, thus, better patient care. 

Nonetheless, the proposed algorithm still has potential for improvement. We aim to include the recognition of serrated lesions, to distinguish normal mucosa from significant inflammatory alterations/diseases, to stratify high-risk lesions into high-grade dysplasia and invasive carcinomas and to identify other neoplasia subtypes. \textcolor{black}{This will enable the prototype to be used upfront in the future.} Further, we would like to leverage the model to be able to evaluate also surgical specimens. Another relevant step will be the merge of our dataset and external ones for training, besides only testing it on external samples. This will enhance its generalisation capabilities and provide a more robust system. Lastly, we intend to measure the “user experience” and feedback from the pathologists, by its gradual implementation in general laboratory routine work. The following goals comprise a more extensive evaluation of the model across more scanner brands and labs. We also want to promote certain mechanisms that would allow for more direct and integrated uncertainty estimation. We have also been looking towards aggregation methods, but, since in the majority of them there is an increased risk of false negatives, we have work to do in that research direction.

\section*{Methods} \label{sec:methods}

In this section, after defining the problem at hand, we introduce the proposed dataset used to train, validate and test the model, the external datasets to evaluate the generalisation capabilities of the model and the pre-processing pipeline. After, we describe in detail the methodology followed to create the deep learning model and to design the experiments. Finally, we also detail the clinical assessment and evaluation of the model. This section also includes a description of the two main bottlenecks that can affect this type of systems. The first is the difficulty of collecting data and having large amounts of data annotated by experts. The second, which becomes apparent only after the first bottleneck is overcame, is the difficulty of scaling these systems as we increase the size of the training data. Without solving the latter problem, it would be impossible to take full advantage of the benefits of collecting large amounts of data. 

\subsection*{Problem definition}

Digitised colorectal cancer histological samples have large dimensions, which are far bigger than the traditional images used in medical or general computer vision problems. Labelling such images is expensive and highly dependent on the availability of expert knowledge. This limits the availability of WSIs, and, in scenarios where these are available, meaningful annotations are usually lacking. On the other hand, it is easier to label the dataset at the slide level. The inclusion of detailed spatial annotations on approximately 10\% of the dataset has been shown to positively impact the performance of deep learning algorithms~\cite{Oliveira2021,neto2022imil4path}. 

\begin{figure}[h!]
    \centering
    \includegraphics[width=0.8\textwidth]{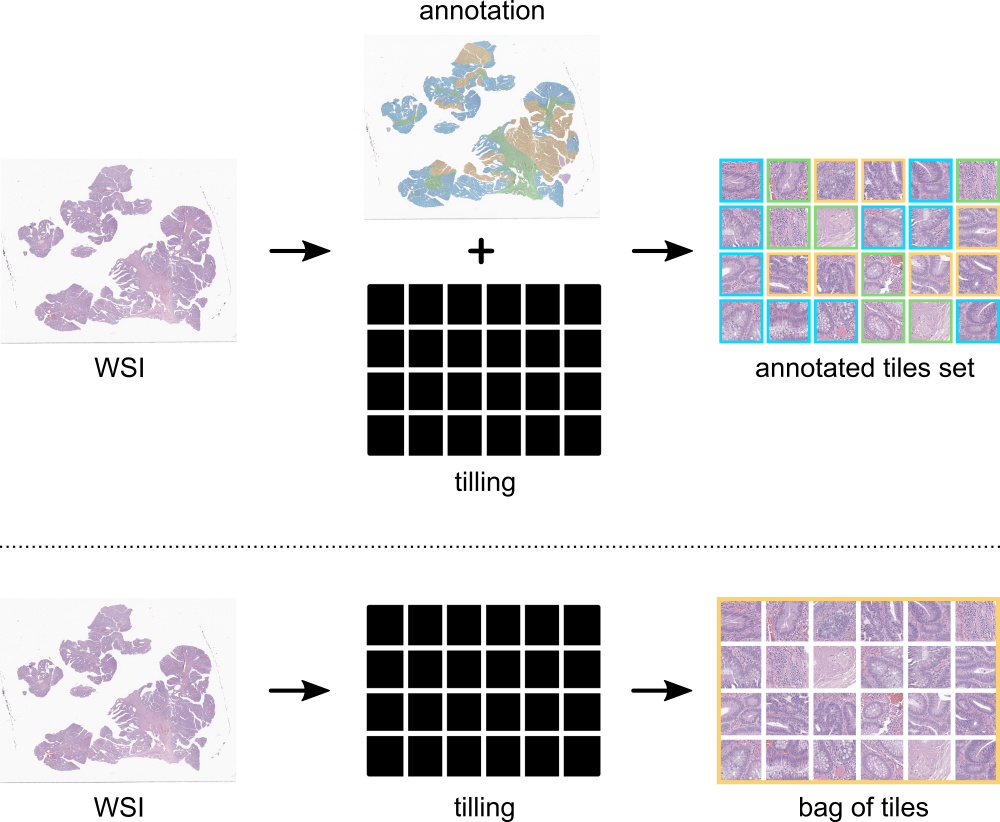}
    \caption{\textbf{Overview of the proposed problem definition}. Problem definition as a fully supervised task (on top), and as a weakly-supervised task (bottom).}
    \label{fig:annota_non_annot}
\end{figure}

To fully leverage the potential of spatial and slide labels, we propose a deep learning pipeline, based on previous approaches~\cite{Oliveira2021,neto2022imil4path}, using mixed supervision. Each slide, ${\mathcal{S}}$  is composed of a set of tiles $\mathcal{T}_{s,n}$, where $s$ represents the index of the slide and $n\in\{1,\cdots, n_s\}$  the tile number. Furthermore, there is an inherent order in the grading used to classify the tiles into one of the $C^{(k)}_{s,n}$ classes, which represents a variation in severity. We define $C^{(k)}_{s}$ and $C^{(k)}_{s,n} \in$ \{``Non-Neoplastic", ``Low-Grade Lesion", ``High-Grade Lesion"\}, and the label of each slide $s$ corresponds to the index $(k)$ of their class. We further define the output of the model as $\hat{y}_{s,n}$ where $\hat{y}_{s,n}(k)$ is the model estimation for $P(C^{(k)}_{s,n}).$ The final prediction of the model is defined as $\hat{C}_{s} \in \{1,..,K\}$ for a slide prediction and $\hat{C}_{s,n} \in \{1,..,K\}$ for a tile prediction, where $K=3$. The latter derives from $\operatorname*{arg\,max}_k P(C^{(k)}_{s,n})$.  For fully supervised learning, only strongly annotated slides are useful, and for those, the class of each tile $C^{(k)}_{s,n}$ is known. The remaining slides are deprived of these detailed labels,  hence, they can only be leveraged by training algorithms with weakly supervision using slide level labels as $C^{(k)}_{s}$. To be used by these algorithms, the weakly annotated slides have only a single label for the entire bag (set) of tiles, as seen in Figure~\ref{fig:annota_non_annot}. Following the order of the labels and the clinical knowledge, we assume that the predicted slide label $\hat{C}_s$ is the most severe case of the tile labels: 

$\hat{C}_s = \max_n \{\hat{C}_{s,n}\}.$ 

In other words, if there is at least one tile classified as containing high-grade dysplasia, then the entire slide that contains the tile is labeled/annotated accordingly. On the other end of the spectrum, if the worst tile is labeled as non-neoplastic, then it is assumed that there is no dysplasia in the entire set of tiles. This is a generalisation of multiple-instance learning (MIL) to an ordinal classification problem, as proposed by Oliveira~\textit{et al.}~\cite{Oliveira2021}.

\subsection*{Datasets}

The spectrum of large-scale CRC/CRS datasets is increasing due to the contributions of several researchers~\cite{neto2024dataset}. Two datasets that have been recently introduced in the literature are the CRS1K~\cite{Oliveira2021} and CRS4K~\cite{neto2022imil4path} datasets from our research group.  Since the latter is an extension of the former with roughly four times more slides, it will be the baseline dataset for the remaining of this document. We further extend this with the CRS10K dataset, which contains 9.26x and 2.36x more slides than CRS1K and CRS4K, respectively. We gathered our data retrospectively from IMP Diagnostics' archive, sequentially selecting all cases that matched the study's diagnostic categories. Thus, we performed consecutive sampling and our cases are a representative sample of the study's population, namely regarding pathology distribution across sex and age in the population. Similarly, the number of tiles is multiplied by a factor of 12.2 and 2.58 (Table~\ref{tab:data_summary}). This volume of slides is translated into an increase in the flexibility to design experiments and infer the robustness of the model. Thus, the inclusion of a test set separated from the validation set is now facilitated. All procedures were in accordance with the ethical standards of the 1964 Helsinki declaration and its later amendments and comparable ethical standards. All data was anonymized and data collection and usage were performed in accordance with the General Data Protection Regulation (GDPR) and national laws and regulations. Ethical approval was waived by the local Ethics Committee of INESC TEC in view of the retrospective nature of the study and all the procedures being performed were part of the routine care.

The set is composed of colorectal biopsies and polypectomies (excluding surgical specimens). CRS10K slides are labelled according to three main categories: non-neoplastic (NNeo), low-grade lesions (LG), and high-grade lesions (HG). The first, contains normal colorectal mucosa, hyperplasia and non-specific inflammation. LG lesions correspond to conventional adenomas with low-grade dysplasia. Finally, HG lesions are composed high-grade dysplasia adenomas (including intramucosal carcinomas) and invasive adenocarcinomas. Slides with suspicion or known history of inflammatory bowel disease, infectious diseases, serrated lesions or other polyp types were not included in the dataset.

The slides were digitised with Leica GT450 WSI scanners, at \textcolor{black}{0.26} µm/pixel at 40$\times$ magnification. The cases were initially seen and classified (labelled) by one of three pathologists. The pathologist revised and classified the slides, and then compared the result with the initial report diagnosis (which served as a second-grader). If there was a match between both, no further steps were taken. In discordant cases, a third pathologist served as a tie-breaker. Roughly 9\% of the dataset (967 slides and over a million tiles) were manually annotated by a pathologist and rechecked by the other, in turn, using the Sedeen Viewer software~\cite{Sedeen}. For complex cases, or when the agreement for a joint decision could not be reached, a third pathologist reevaluated the annotation.

The CRS10K dataset was divided into train, validation and test sets. The training set includes all the strongly annotated slides, for fully-supervised learning, and a random selection of weakly-annotated samples. The validation set, on the other hand, consists of only weakly-annotated slides. Finally, the test set was selected from the new data added to extend the previous datasets and only includes weakly-annotated slides. Thus, it is completely separated from the training and validation sets of previous works. The test set, is publicly available, so that future research can directly compare their results and use this set as a benchmark. 

\begin{table} [h!]
\caption{Dataset summary, with the number of slides (annotated samples are detailed in parentheses) and tiles distributed by class for all the datasets used in this study.}
\label{tab:data_summary}
\newcolumntype{C}{>{\centering\arraybackslash}X}
\scriptsize
\begin{tabularx}{\textwidth}{llcccc}
\toprule
& & NNeo & LG & HG & Total\\\midrule
& \# slides & 300 (6) & 552 (35) & 281 (59) & 1133 (100)\\
CRS1K dataset~\cite{Oliveira2021} & \# annotated tiles & 49,640 & 77,946 & 83,649 & 211,235\\
& \# non-annotated tiles & - & - & - & 1,111,361 \\
\midrule
& \# slides & 663 (12) & 2394 (207) & 1376 (181) & 4433 (400)\\
CRS4K dataset~\cite{neto2022imil4path} & \# annotated tiles & 145,898 & 196,116 & 163,603 & 505,617\\
& \# non-annotated tiles & - & - & - & 5,265,362\\
\midrule
& \# slides & 1740 (12) & 5387 (534) & 3369 (421) & 10,496 (967)\\
CRS10K dataset & \# annotated tiles & 338,979 & 371,587 & 341,268 & 1,051,834\\
& \# non-annotated tiles & - & - & - & 13,571,871\\
\midrule
CRS Prototype & \# slides & 28 &  44 &  28 &  100\\
& \# non-annotated tiles & - & - & - & 244,160\\
\midrule
PAIP~\cite{PAIP}& \# slides & - &  - &  100 &  100\\
& \# non-annotated tiles & - & - & - & 97,392\\
\midrule
TCGA~\cite{Clark2013} & \# slides &  1 &  1 & 230  &  232\\
& \# non-annotated tiles & - & - & - & 1,568,584\\
\bottomrule
\end{tabularx}
\end{table}

Of note, when collected from routine archives, the slides can be digitised with duplicated tissue areas. Hence, the workflow for the automatic diagnostic also included an automatic fragment detection and counting system, to avoid repeated and lower quality fragments ~\cite{tome2022automaticcount}. 

Furthermore, as detailed in the following sections, this work comprises the development of a fully-functional prototype to be used in clinical practice. Leveraging this prototype, it was possible to further collect a new set with 100 slides. It differs from the CRS10K dataset, as these cases were actively collected from the current year's routine exams. We argue that this might better reflect the real-world data distribution. Hence, we introduce this set as a distinct dataset to evaluate the robustness of the presented methodology. Differently from the datasets discussed below, the CRS Prototype dataset has a more balanced distribution of the slide labels. Although useful, using the fragment counting and selection algorithm for the evaluation could potentiate the propagation of errors from one system to another. Thus, in this evaluation, we did not use the fragment selection algorithm, and as shown in Table~\ref{tab:data_summary}, the number of tiles per slide doubles when compared to CRS10K, which had its fragments carefully selected.

To evaluate the domain generalisation of the proposed approach, two external datasets, publicly available, were used. The first dataset is composed of samples of the TCGA-COAD~\cite{Kirk2016} and TCGA-READ~\cite{Kirk2016b} collections from The Cancer Imaging Archive~\cite{Clark2013}, which are composed in general by resection samples (in contrast to our dataset, composed only of biopsies and polypectomies). Samples containing pen markers, large air bubbles over tissue, tissue folds, and other artefacts affecting large areas of the slide were excluded. The final selection includeded 232 whole-slide images reviewed and validated by the same pathologists that reviewed our in-house datasets. 230 of those samples were diagnosed as high-grade lesions, whereas the remaining two have been diagnosed as low-grade and non-neoplastic. For this dataset, the specific model of the scanner used to digitise the images is unknown, but the file type (".svs") matches the file type of the training data.  The second external dataset contains 100H\&E slides from the Pathology AI Platform~\cite{PAIP} colorectal cohort. All included cases had a more superficial sampling of the lesions, better comparing with our datasets. All the WSIs in this dataset were digitised with an Aperio AT2 at 20X magnification. Finally, the pathologists' team followed the same guidelines to review and validate all the WSI, which were all classified as high-grade lesions. It is interesting to note that while the PAIP contains significantly fewer tiles per slide, around 973, than the CRS10K dataset, around 1293, the TCGA dataset shows the largest amount of tissue per slide with an average of 6761 tiles as seen in Table~\ref{tab:data_summary}.

\subsection*{Data pre-processing}

\begin{figure}[h!]
    \centering
    \includegraphics[width=0.9\linewidth]{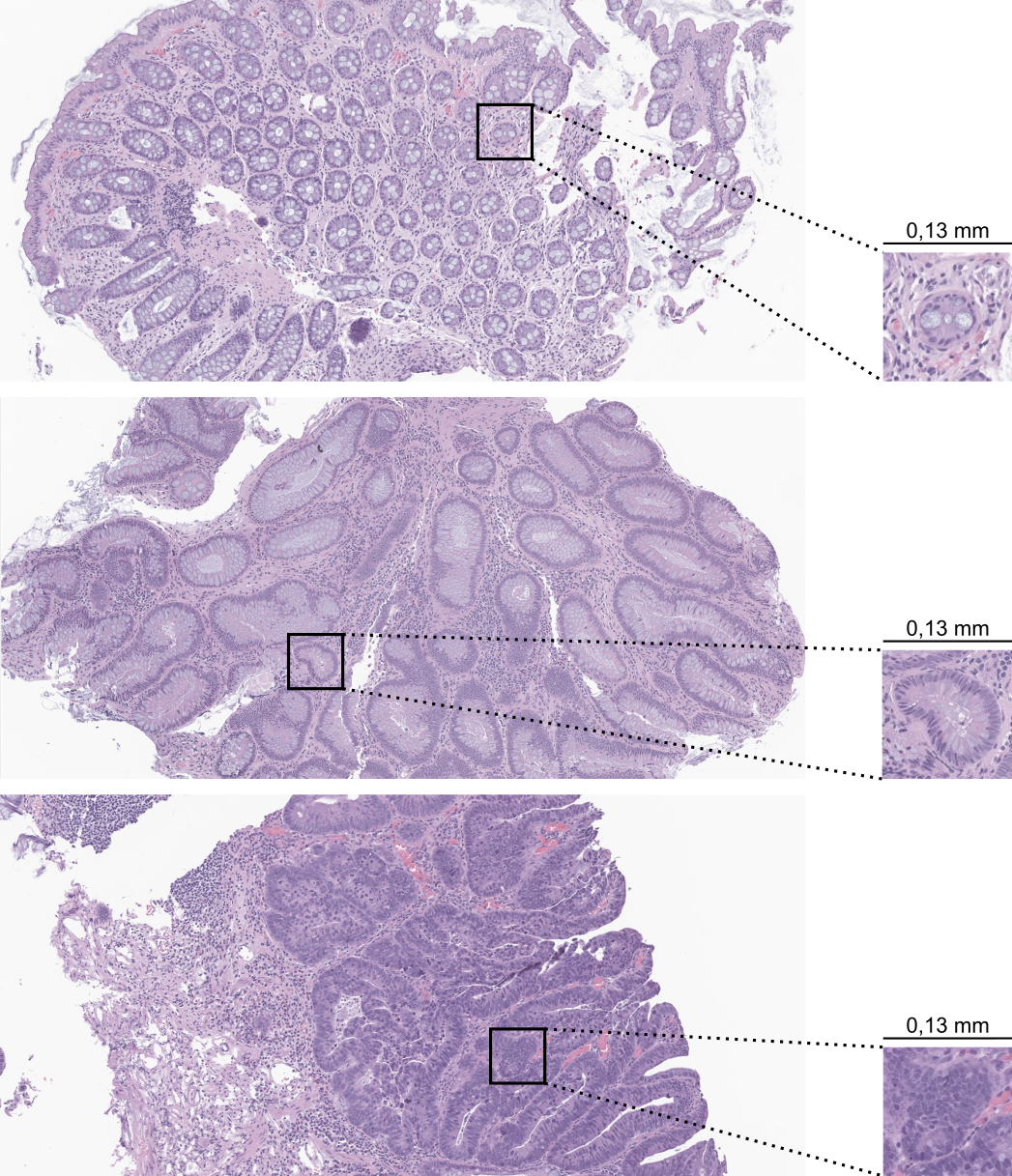}
    \caption{\textbf{Examples of regions and a sample tile with $512\times512$ pixels (40$\times$ magnification)}.  The represented classes are: non-neoplastic (on top), low-grade dysplasia (on the middle) and high-grade dysplasia (on the bottom) with a width and height of 0.13 milimeters (mm) each.}
    \label{fig:tiles}
\end{figure}

H\&E slides are composed of two distinct elements, white background and colourful tissue. Since the former is not meaningful for the diagnostic, the pre-processing of these slides incorporates an automatic tissue segmentation with Otsu’s thresholding~\cite{Otsu1979} on the saturation (S) channel of the HSV colour space, resulting in a separation between the tissue regions and the background. The result of this step, which receives as input a $32\times$ downsampled slide, is the mask used for the following steps. Leveraging this previous output, tiles with a dimension of \mbox{$512\times512$ pixels} (Figure~\ref{fig:tiles}) were extracted from the original slide (without any downsampling) at its maximum magnification (40$\times$), if they did not include any portion of background (i.e. a 100\% tissue threshold was used). Following previous experiments in the literature, our empirical assessment, and the confirmation that smaller tiles would significantly increase the number of tiles and the complexity of the task, $512\times512$ was chosen as the tile size. Moreover, it is believed that $512\times512$ is the smallest tile size that still incorporates enough information to make a good diagnostic with the possibility of visually explaining the decision~\cite{Oliveira2021}. The selected threshold of 100\% further reduces the number of tiles by not including the tissue at the edges and decreases the complexity of the task, since the model does not see the background at any moment. Due to tissue variations in different images, there is also a different number of tiles extracted per image.

\subsection*{Methodology}

The massive size of images, which translates to thousands of tiles per image, allied to a large number of samples in the CRS10K dataset, bottlenecks the training of weakly-supervised models based on multiple instance learning (MIL). Hence, in this document, we propose a mix-supervision approach with self-contained tile sampling to diagnose colorectal cancer samples from WSIs. This subsection comprises the methodology, which includes supervised training, sampling and weakly-supervised learning.

\begin{figure}[h!]
    \centering
    \includegraphics[width=\textwidth]{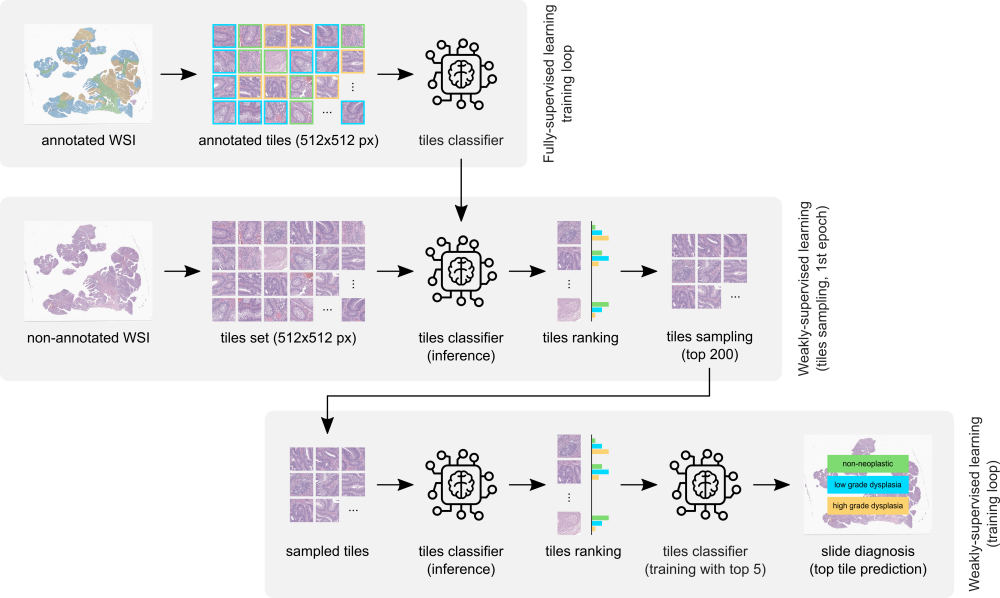}
    \caption{\textbf{Scheme for the proposed mixed precision workflow.} Overall scheme of the proposed methodology containing the mix-supervision framework that is responsible for diagnosing colorectal samples from WSI. The top layer consists of the fully-supervised stage, the middle layer consists of the sampling strategy and the bottom layer represents the weakly supervised training stage.  Tile sizes are in pixels (px).}
    \label{fig:overall_structure}
\end{figure}

\subsubsection*{Supervised Training}

As mentioned in previous sections, spatial annotations are rare in large quantities. However, these include domain information, given by the expert annotator, concerning the most meaningful areas and what are the most and less severe tiles. Thus, they can facilitate the initial optimisation of a deep neural network. As shown in the literature, there has been some research on the impact of starting the training with a few iterations of fully-supervised training~\cite{Oliveira2021,bovzivc2021mixed}. We further explore this in three different ways. First, we have 967 annotated slides resulting in more than one million annotated tiles for supervised training. Secondly, attending to the size of our dataset and the need for a stronger initial supervised training, the models are trained for 50 epochs, and their performance was monitored over specific checkpoint epochs. Finally, we explore this pre-trained model as the main tool to sample useful tiles for the weakly-supervised task.

\subsubsection*{Tile Sampling}
Our scenario presents a particularly difficult condition for scaling the training data. First, let's consider the structure of the data, which consists of, on average, more than one thousand tiles per slide. If we carefully analyse Table~\ref{tab:data_summary}, we can see that the CRS10K dataset and the CRS Prototype contain, when combined, $\approx$13.8 million tiles. These tiles come after preprocessing, as such, tiles containing background are not included. If we further analyse this data, and considering that each tile is of dimension $512\times512\times3$, then we have $\approx$3.6 trillion pixels per colour channel, or $\approx$10.9 trillion pixels in total. Li~\textit{et al.}\cite{li2023scaling}\ described the difficulties of processing 399 WSI in a single GPU. With the following strategy we processed all the 10928 WSI described in Table~\ref{tab:data_summary} utilising a single GPU.

Within the set of tiles from a slide, some tiles provide meaningful value for the prediction, and others do not add extra information. In other words, for the CRS10K dataset, the extensive, time and energy-consuming process of going through 13 million tiles every epoch can be avoided, and, as result, these models can be trained for more epochs. Nowadays, there is an increasing concern regarding energy and electricity consumption. Thus, these concerns, together with the sustainability goals, further support the importance of more efficient training processes.

{Let $\mathcal{T}$ be the original set of tiles, and $\mathcal{T}_s$ be the original set of tiles from the slide $s$, the former is composed by a union of the latter of all the slides (Eq.~\ref{equation:union_of_tile_set}). We propose to map $\mathcal{T}$ to a smaller set of tiles $\mathcal{M}$ without affecting the overall performance and behaviour of the trained algorithm.}

\begin{equation}
    \label{equation:union_of_tile_set}
    \mathcal{T} = \bigcup\limits_{s=1}^{S} \mathcal{T}_{s}
\end{equation}

The model trained in a fully supervised task, previously described, provides a good estimation of the utility of each tile. Hence, we utilise the function  ($\Phi$) learned by the model to compute the predicted severity of each tile. In other words, $\Phi$ represents the supervised model already trained. 
We select M tiles per slide (M=200 in our experimental setup) utilising a Top-k function (with k set to 200) to be retained for the weakly-supervised training. As indicated by the results presented in the following sections, the value of M was selected in accordance with a trade-off between information lost and training time. This is formalised in Eq.~\ref{equation:sampled_set}.

\begin{equation}
    \label{equation:sampled_set}
    \mathcal{M}_s = \textrm{Top-k}(\Phi(\mathcal{T}_s))
\end{equation}

For instance, in the CRS10K dataset, the total number of tiles after sampling would be at most 2,099,200, which represents a reduction of 6.46$\times$ when compared to the total number of slides. Despite this upper bound on the number of tiles, there are WSI samples that contain less than M tiles, and as such, they remain unsampled and the actual total number of tiles after sampling is potentially lower. During the evaluation and test time, there is no sampling.

\subsubsection*{Weakly-Supervised Learning}

The weakly-supervised learning approach designed for our methodology follows the same principles of recent work~\cite{neto2022imil4path}. It is divided into two distinct stages, tile severity analysis and training. The former utilises the pre-trained model to evaluate the severity of every tile in a set of tiles. In the first epoch, $\mathcal{T}$, the set of all the tiles in the complete dataset is used. This is possible since the model used to assess the severity in this epoch is the same one used for sampling. Hence, both tasks are integrated with the initial epoch. The following epochs utilise the sampled tile set $\mathcal{M}$ instead of the original set. In other words, the bags (i.e. the representation of the slides) are all truncated to size 200. This overall structure is represented in Figure~\ref{fig:overall_structure}. Moreover, the weakly-supervised approach leverages only slide labels.

{The link between both stages is guided by a slide-wise tile ranking approach based on the expected severity as proposed in \cite{Oliveira2021}. For tile $\mathcal{T}_{s,n}$, the expected severity is defined as}



\begin{equation}
    \label{eqExpectedV}
\mathbb{E}(\hat{C}_{s,n}) = \sum_{i=1}^K i \times \hat{y}_{s,n}(i)
\end{equation}


\noindent where $\hat{y}_{s,n}(i)$ is a random vector of size $K$, which represents the $P(C^{(i)}_{s,n})$ for the tile $n$ of the slide $s$.
After this analysis, the five most severe tiles are selected from each bag of 200 tiles for training. The number of selected tiles was chosen in accordance with previous studies~\cite{neto2022imil4path}. These five tiles per bag are used to train the proposed model for one more epoch. An epoch is composed of both stages, which means that the tiles used for training vary across epochs. The slide label is used as the ground truth of all five tiles of that same slide used for network optimisation. For validation and evaluation, only the most severe tile is used for diagnostics. Although it might lead to an increase in false positives, it shall significantly reduce false negatives. Furthermore, we argue that increasing the variability and quantity of data available leads to a better balance between the reduction of these two types of errors.

\subsection*{Reject option}

Automatic systems designed to assist pathologists should be high-performing and achieve outstanding values in evaluation metrics. However, it is equally important for these systems to recognize their limitations and defer to expert pathologists in challenging cases. Recognizing the importance of this feature, we introduce a reject option to our model. Pathologists can further tune the expected rate of rejection and the performance on a set of metrics to better suit the model to their needs.  

The adopted strategy creates the possibility to reject a sample based on the predicted probability of the predicted label. Then, the desired rejection rate is calculated from the percentiles of all confidence values. This approach magnifies the innate capabilities of deep learning systems to be used as a second/third opinion system.

\subsection*{Confidence Interval}

\textcolor{black}{In order to quantify the uncertainty of a result, it is common to compute the 95 percent confidence interval. In this way, two different models can be easily understood and compared based on the overlap of their confidence intervals. The standard approach to calculating these intervals requires several runs of a single experiment. As we increase the number of runs, our interval becomes narrower. However, this procedure is impractical for the computationally intensive experiments presented in this document. Hence, we use an independent test set to approximate the confidence interval as a Gaussian function~\cite{raschka2018model}. To do so, we compute the standard error ($SE$) of an evaluation metric $m$, which is dependent on the number of samples ($n$), as seen in Equation~\ref{eq:se}.}

\begin{equation}
\label{eq:se}
    SE = \sqrt{\frac{1}{n} \times m \times (1-m)}
\end{equation}

\textcolor{black}{For the SE computation to be mathematically correct, the metric $m$ must originate from a set of Bernoulli trials. In other words, if each prediction is considered a Bernoulli trial, then the metric should classify them as correct or incorrect. The number of correct samples is then given by a Binomial distribution $X \sim (n,p)$, where $p$ is the probability of correctly predicting a label, and $n$ is the number of samples. For instance, the accuracy is a metric that fits all these constraints.}

\textcolor{black}{Following the definition and the properties of a Normal distribution, we compute the number of standard deviations ($z$), known as a standard score, that can be translated to the desired confidence ($c$) set to 95\% of the area under a normal distribution. This is a well-studied value, which is approximately $z\approx1.96$. This value $z$ is then used to calculate the confidence interval, calculated as the product of $z$ and $SE$ as seen in Equation~\ref{eq:se_z}.}

\begin{equation}
\label{eq:se_z}
    M \pm z * \sqrt{\frac{1}{n} \times m \times (1-m)}
\end{equation}

\textcolor{black}{To infer the statistical significance of the different performance of different classifiers the McNemar's test~\cite{mcnemar1947note} was used. These statistical tests have been used in the literature for comparison of independent systems and there are several variations~\cite{demvsar2006statistical}. For the McNemar's test the classifiers must be compared in pairs. For each pair, it is necessary to build a contigency table containing four entries: a) samples misclassified by both; b) samples misclassified only by the second classifier; c) samples misclassified only by the first classifier; d) samples correctly classified by both. The null hypothesis of this test states that the second (b) and third (c) entries have equal probability. $\mathcal{X}^2$ corrected for continuity~\cite{edwards1948note}is calculated as follows:}

\begin{equation}
\label{eq:x2_mc}
    \mathcal{X}^2 = \frac{(|b-c|-1)^2}{b+c}
\end{equation}

\textcolor{black}{Using a significance value of 0.95, we can reject the null hypothesis if $\mathcal{X}^2 > 3.841$, which corresponds to the area between 0.05 and  $+\infty$ for a Chi-Squared distribution with 1 degree of freedom.}

\subsection*{Label correction}

\textcolor{black}{The complex process of labelling thousands of whole-slide images with colorectal cancer diagnostic grades is a task of increased difficulty. It should also be noted and taken into account that grading colorectal dysplasia is hurdled by considerable subjectivity, so it is to be expected that some borderline cases will be classified by some pathologists as low-grade and others as high-grade.  Moreover, as the number of cases increases, it becomes increasingly difficult to maintain perfect criteria and avoid mislabelling. For this reason, we have extended the analysis of the model's performance to understand its errors and its capability to detect mislabelled slides.}

\textcolor{black}{After training the proposed model, it was evaluated on the test data. Following this evaluation, we identified the misclassified slides and conducted a second round of labelling. These cases were all blindly reviewed by two pathologists, and discordant cases from the initial ground truth were discussed and classified by both pathologists (and in case of doubt/complexity, a third pathologist was also consulted). We tried to maintain similar criteria between the graders and always followed the same guidelines. These new labels were used to rectify the performance of all the algorithms evaluated in the test set. We argue that the information regarding the strength/confidence of predictions of a model used as a second opinion is of utter importance. A correct integration of this feature can be shown as extremely insightful for the pathologists using the developed tool.}

\subsection*{Experimental setup}
\textcolor{black}{For our experimental setup, we divide our data into training and validation sets. Besides, we further evaluate the performance of the former in our test set composed of slides never seen by any of the methods presented or in the literature. Following the split of these three sets, we have 8587, 1009 and 900 stratified non-overlapping samples in the training, validation and test set, respectively.}

\textcolor{black}{In an attempt to also contribute to reproducible research, the training of all the versions of the proposed algorithm uses the deterministic constraints available on PyTorch. The usage of deterministic constraints implies a trade-off between performance, either in terms of algorithmic efficiency or on its predictive power, and the complement with reproducible research guidelines. As such, due to the current progress in the field, we have chosen to comply with the reproducible research guidelines.}

\textcolor{black}{All the trained backbone networks were ResNet-34~\cite{He2016} with ImageNet weights. PyTorch was used to train these networks with the Adaptive Moment Estimation (Adam)~\cite{kingma2015adam} optimiser, a learning rate of $6\times10^{-6}$ and a weight decay of $3\times10^{-4}$. The training batch size was set to 32 for both fully and weakly supervised training, while the test and inference batch size was 256. The performance of the model was verified on the validation set used for model selection in terms of the best accuracies and quadratic weighted kappa (QWK). The training was accelerated by an Nvidia Tesla V100 (32GB) GPU for 50 epochs of both weakly and fully supervised learning. In addition to the proposed methodology, we extended our experiments to include the aggregation approach proposed by Neto~\textit{et al.}~\cite{neto2022imil4path} on top of our best-performing method. This strategy does not consider spatial location or context of the tiles, instead it select the seven most severe tiles, concatenates the output of the last convolutional layer for each of those tiles and feeds it to a multi-layer perceptron.}

\textcolor{black}{The number of epochs for training the fully- and weakly-supervised models was selected as follows: the fully supervised model was evaluated at every ten epochs for its performance on a weakly supervised scenario (using the non annotated samples), when its performance was stable the training was stopped; for the weakly-supervised model, several experiments were conducted on smaller versions of the training set and validated on the validation set. For the latter, besides training for 50 epochs, the best weights (with respect to the performance on the validation set) were selected.}

\subsection*{Prototype and Interpretability Assessment}

\begin{figure}[h!]
    \centering
    \includegraphics[width=\textwidth]{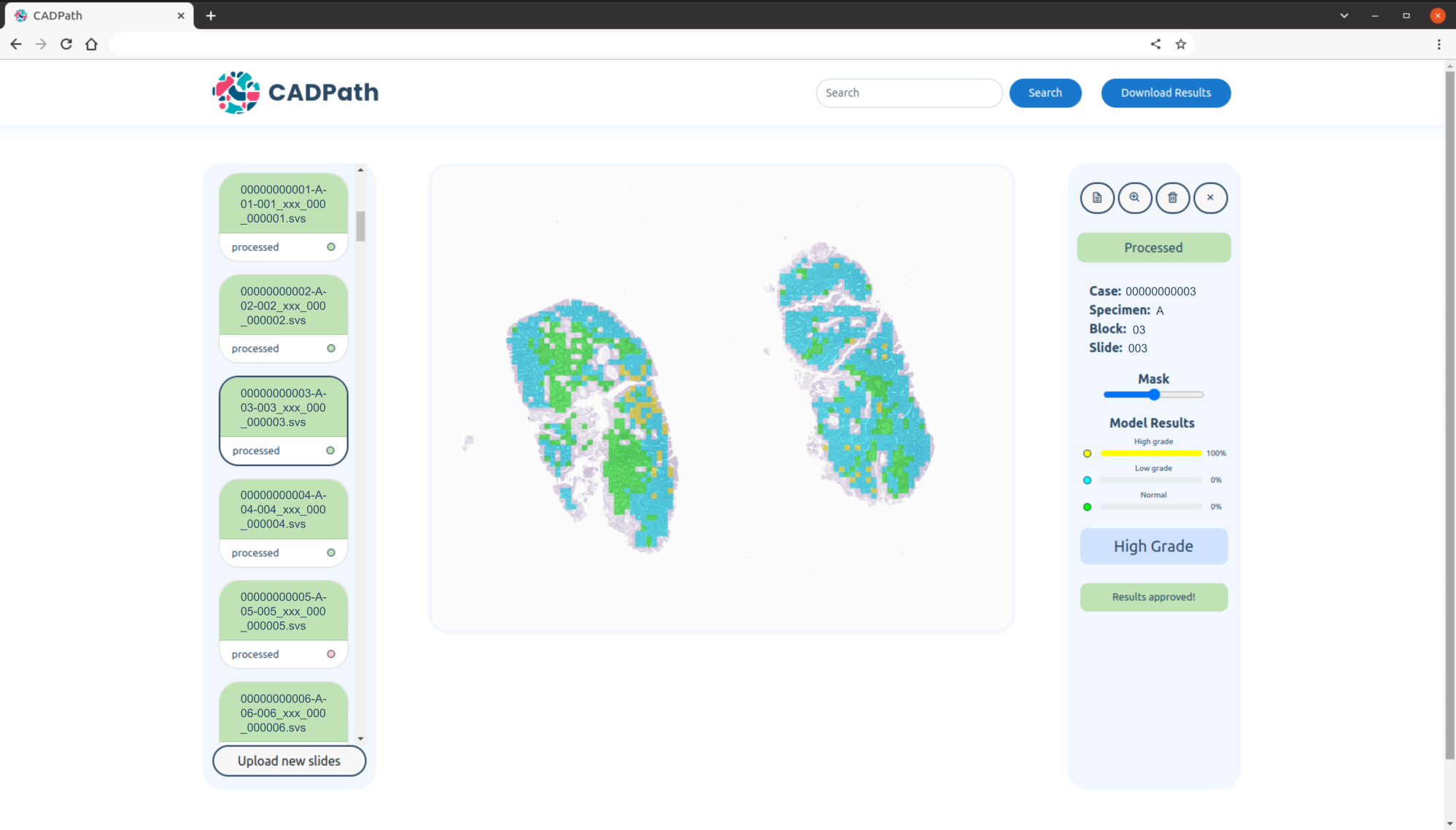}
    \caption{\textbf{Main view of the CAD system prototype for CRS:} Slide identification, confidence per class, diagnostic, mask overlay controller, results download as csv and slide search are some of the features visible. Slide identification is anonymised.}
    \label{fig:prototype_base}
\end{figure}

\begin{figure}[h!]
    \centering
    \includegraphics[width=\textwidth]{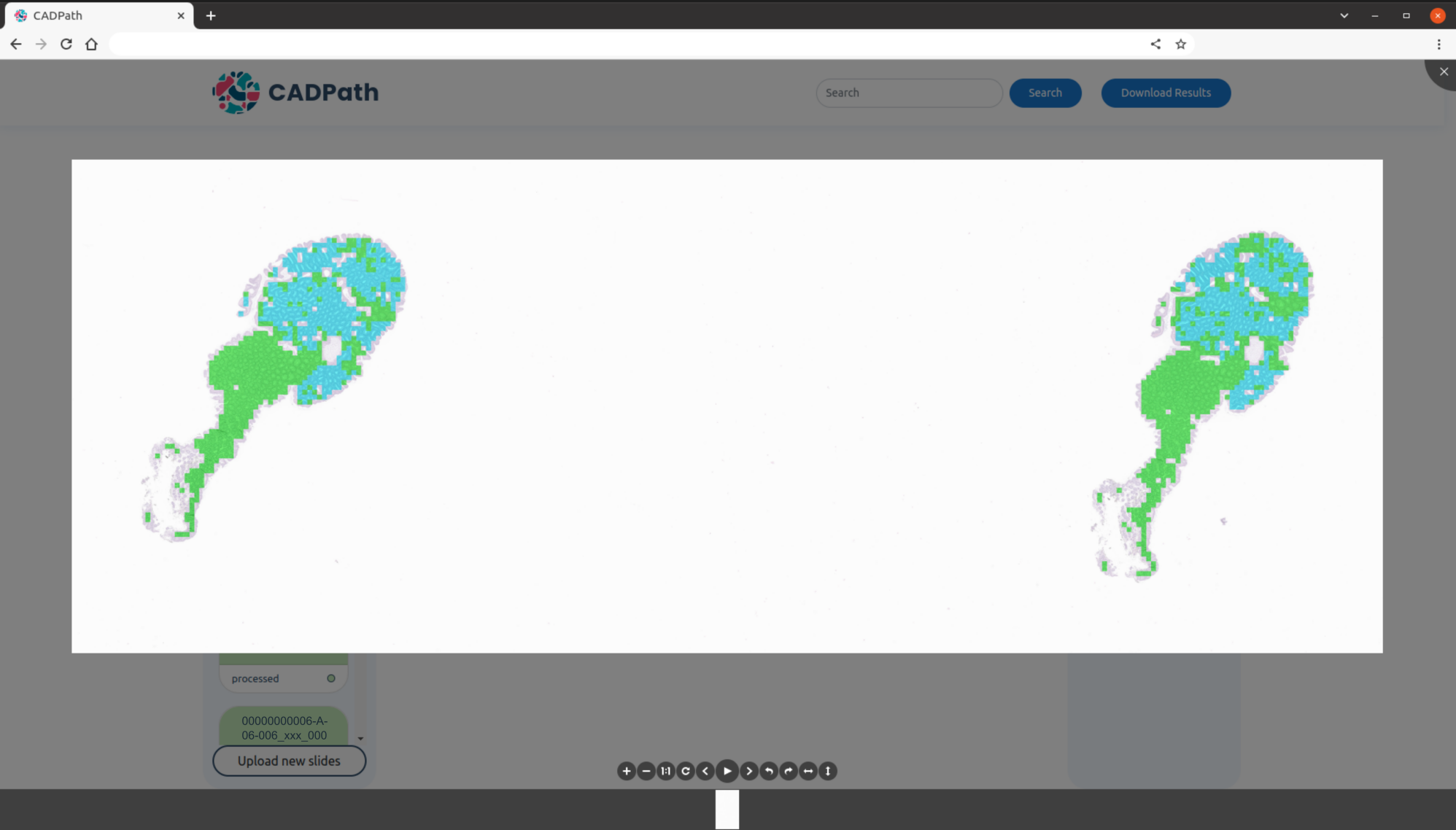}
    \caption{\textbf{Zoomed view of a slide from the CAD system prototype.} Further includes the predictions map with a small overlay threshold. Slide identification is anonymised.}
    \label{fig:prototype_map}
\end{figure}

The proposed algorithm was integrated into a fully functional prototype to enable its use and validation in a real clinical workflow. This system was developed as a server-side web application that can be accessed by any pathologist in the lab. The system supports the evaluation of either a single slide or a batch of slides simultaneously and in real time. It also caches the most recent results, allowing re-evaluation without the need to re-upload slides. In addition to displaying the slide diagnosis, and confidence level for each class, a visual explanation map is also retrieved, to draw the pathologist’s attention to key tissue areas within each slide (all seen in Figure~\ref{fig:prototype_base}). The opaqueness of the map can be set to different thresholds, allowing the pathologist to control its overlay over the tissue. An example of the zoomed version of a slide with lower overlay of the map is shown in Figure~\ref{fig:prototype_map}.

\begin{figure}[h!]
    \centering
    \includegraphics[width=\textwidth]{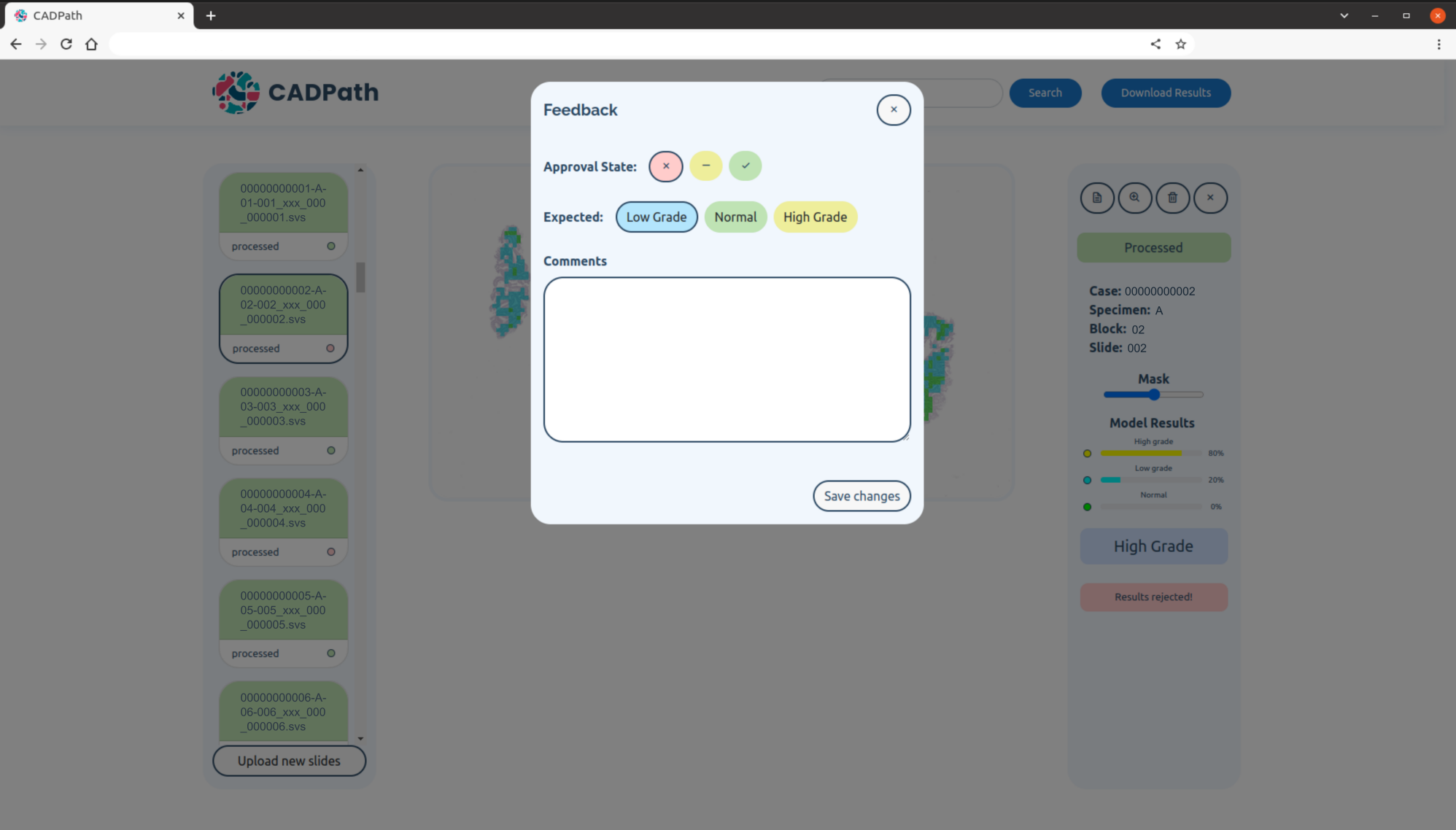}
    \caption{\textbf{CAD system prototype report tool:} the user can report if the result is either correct, wrong or inconclusive and leave a comment for each case individually. Slide identification is anonymised.}
    \label{fig:prototype_feedback}
\end{figure}

Furthermore, the prototype also allows user feedback where the user can accept/reject a result and provide a justification (Figure~\ref{fig:prototype_feedback}), an important feature for software updates, research development and possible active learning frameworks that can be developed in the future. These results can be downloaded with the corrected labels to allow for further retraining of the model.

There are several advantages to developing such a system as a server-side web application. First, it does not require any specific installation or dedicated local storage in the user's device. Secondly, it can be accessed at the same time by several pathologists from different locations, allowing for a quick review of a case by multiple pathologists without data transference. Moreover, the lack of local storage of clinical data increases the privacy of patient data, which can only be accessed through a highly encrypted virtual private network (VPN).  Finally, all the processing can be moved to an efficient graphics processing unit (GPU), thus reducing the processing time by several orders of magnitude. Similar behaviour on a local machine would require the installation of dedicated GPUs in the pathologists' personal devices. This platform is the first Pathology prototype for colorectal diagnosis developed in Portugal, and, as far as we know, one of the pioneers in the world. Its design was also carefully thought to be aligned with the needs of the pathologists.

\section*{Data Availability}

A large portion of the dataset has been publicly released~\cite{neto2024dataset} and is identified by the following DOI: \url{https://doi.org/10.25747/fb1q-j507}. This data composed of WSI and respective labels has been released under CC BY-NC. This release is part of the efforts of IMP Diagnostics and INESC TEC to advance science and share knowledge. 

\section*{Acknowledgements}

This work is financed by National Funds through the Portuguese funding agency, FCT - Fundação para a Ciência e a Tecnologia, within project LA/P/0063/2020 and the PhD Grants SFRH/BD/139108/2018 and 2021.06872.BD. Additionally, data collection, infrastructure and resources have been supported by the European Regional Development Fund (ERDF), through the Operational Programme for Competitiveness and Internationalisation (COMPETE 2020) Programme, within the project POCI-01-0247-FEDER-045413 and by IMP Diagnostics.

\section*{Author Contribution}
P.C.N., S.P.O. and D.M. contributed equally to this work; P.C.N., S.P.O. and J.S.C. designed the experiments; P.C.N. and S.P.O. conducted the experiments and the analysis of the results; J.M, L.R. and S.G. prepared the histopathological samples; J.F., D.O. and D.M. collected, reviewed and annotated the histopathological cases; P.C.N., S.P.O. and D.M. wrote the manuscript; A.M. and J.M. performed data curation and project management; I.M.P., I.Z. and S.R. clinically supervised the project; J.S.C., I.Z. and S.R. technically supervised the project. All authors have read and agreed to the published version of the manuscript.

\section*{Competing Interests}

The authors declare the following competing interests: D.M. and D.O. are employees at IMP Diagnostics and I.M.P. is an owner at IMP Diagnostics.

\bibliographystyle{elsarticle-num} 
\bibliography{references.bib}

\listoffigures

\begin{enumerate}
    \item Figure 1 - \textbf{Tile sampling impact on information loss}: percentage of tiles not selected due to sampling with different thresholds, over the first four inference epochs. The blue bar represents a sampling strategy that retains 200 tiles per slide, the orange bar is for a strategy that retains 100 tiles, the green bar represents a strategy that retains 75 tiles and finally the strategy represented by the red line retains 50 tiles per slide. 
    \item Figure 2 - \textbf{Precision-recall curve on the on the CRS10K test set:} For the three distinct models, we have calculated the Precision-recall curve on this dataset. Includes an indication of the F1-Score for each of the different models. The blue line represents the curve of Our method when trained on CRS10K, while the orange line shows the same method when trained on CRS4K. The green line is the curve of iMIL4Path.
    \item Figure 3 - \textbf{Confidence analysis for correct and incorrect predictions on the CRS10K test set:} Kernel density estimation of the confidences of correct and incorrect predictions performed on the three-class classification problem by three distinct models on the CRS10K test set. The plots represent, from left to right, the proposed method trained on CRS10K, the proposed method trained on CRS4K and iMIL4Path. In each plot, the blue line defines the density function of the correct samples and the blue dashed line the mean confidence of those samples. On the other hand, the orange solid and dashed lines represent the same for incorrect predictions.
    \item Figure 4 - \textbf{Confidence analysis for correct and incorrect predictions on the Prototype set:} Kernel density estimation of the confidences of correct and incorrect predictions performed on the three-class classification problem by three distinct models on the prototype set. The plots represent, from left to right, the proposed method trained on CRS10K, the proposed method trained on CRS4K and iMIL4Path. In each plot, the blue line defines the density function of the correct samples and the blue dashed line the mean confidence of those samples. On the other hand, the orange solid and dashed lines represent the same for incorrect predictions.
    \item Figure 5 - \textbf{Accuracy-vs-Rejection-rate for the models evaluated on the CRS10K test set}. Relation between the accuracy and the percentage of samples not classified by the model. Both axes are in percentage. The blue line represents Our method when trained on CRS10K, while the orange line shows the same method when trained on CRS4K. The green line is for iMIL4Path
    \item Figure 6 - \textbf{Overview of the proposed problem definition}. Problem definition as a fully supervised task (on top), and as a weakly-supervised task (bottom).
    \item Figure 7 - \textbf{Examples of regions and a sample tile with $512\times512$ pixels (40$\times$ magnification).}  The represented classes are: non-neoplastic (on top), low-grade dysplasia (on the middle) and high-grade dysplasia (on the bottom) with a width and height of 0.13 milimeters (mm) each.
    \item Figure 8 - \textbf{Scheme for the proposed mixed precision workflow.} Overall scheme of the proposed methodology containing the mix-supervision framework that is responsible for diagnosing colorectal samples from WSI. The top layer consists of the fully-supervised stage, the middle layer consists of the sampling strategy and the bottom layer represents the weakly supervised training stage.  Tile sizes are in pixels (px).
    \item Figure 9 - \textbf{Main view of the CAD system prototype for CRS:} Slide identification, confidence per class, diagnostic, mask overlay controller, results download as csv and slide search are some of the features visible. Slide identification is anonymised
    \item Figure 10 - \textbf{Zoomed view of a slide from the CAD system prototype.} Further includes the predictions map with a small overlay threshold. Slide identification is anonymised.
    \item Figure 11 - \textbf{CAD system prototype report tool:} the user can report if the result is either correct, wrong or inconclusive and leave a comment for each case individually. Slide identification is anonymised.
\end{enumerate}

\end{document}


\begin{frontmatter}



\title{\textbf{Supplementary material for: }\\An interpretable machine learning system for colorectal cancer diagnosis from pathology slides}

\author[inesc,feup,fn1]{Pedro C. Neto}
\author[imp,icbas,ipo1,fn1]{Diana Montezuma}
\author[inesc,feup,fn1]{Sara P. Oliveira}
\author[imp]{Domingos Oliveira}
\author[ipo2]{João Fraga}
\author[imp]{Ana Monteiro}
\author[imp]{João Monteiro}
\author[imp]{Liliana Ribeiro}
\author[imp]{Sofia Gonçalves}
\author[bern]{Stefan Reinhard}
\author[bern]{Inti Zlobec}
\author[imp]{Isabel M. Pinto}  
\author[inesc,feup]{Jaime S. Cardoso}

\fntext[fn1]{These authors contributed equally.}

\affiliation[inesc]{organization={Institute for Systems and Computer Engineering, Technology and Science (INESC TEC)},
             addressline={R. Dr. Roberto Frias},
             city={Porto},
             postcode={4200-465},
             state={Porto},
             country={Portugal}}

 \affiliation[feup]{organization={Faculty of Engineering, University of Porto (FEUP)},
             addressline={R. Dr. Roberto Frias},
             city={Porto},
             postcode={4200-465},
             state={Porto},
             country={Portugal}}
 \affiliation[imp]{organization={IMP Diagnostics},
             addressline={Praça do Bom Sucesso, 61, sala 809},
            city={Porto},
            postcode={4150-146},
            state={Porto},
            country={Portugal}}
\affiliation[ipo1]{organization={Cancer Biology and Epigenetics Group, IPO-Porto},
            addressline={R. Dr. António Bernardino de Almeida 865},
            city={Porto},
            postcode={4200-072},
            state={Porto},
            country={Portugal}}
\affiliation[ipo2]{organization={Department of Pathology, IPO-Porto},
            addressline={R. Dr. António Bernardino de Almeida 865},
            city={Porto},
            postcode={4200-072},
            state={Porto},
            country={Portugal}}
\affiliation[icbas]{organization={School of Medicine and Biomedical Sciences, University of Porto (ICBAS)},
            addressline={R. Jorge de Viterbo Ferreira 228},
            city={Porto},
            postcode={4050-313},
            state={Porto},
            country={Portugal}}

\affiliation[bern]{organization={Institute of Pathology, University of Bern},
            addressline={Uni Bern, Murtenstrasse 31},
            city={Bern},
            postcode={3008},
            state={Bern},
            country={Switzerland}}






\begin{keyword}
Clinical Prototype \sep Colorectal Cancer \sep Interpretable Artificial Intelligence	\sep Deep Learning \sep Whole-Slide Images

\end{keyword}

\end{frontmatter}

\section*{Supplementary Results}

\textcolor{black}{Supplementary Tables~\ref{tab:test_other_metrics},~\ref{tab:prototype_other_metrics} and \ref{tab:all_other_metrics} show a consistent advantage of the proposed model when trained on the CRS10K dataset over the other models. This metrics provide more insights on the performance of the model and corroborate the development of the CRS10K model. }

\textcolor{black}{On Supplementary Table~\ref{tab:prototype_other_metrics} it is visible that despite having the same accuracy, iMIL4Path provides less separable confidences when compared with our method trained on CRS10K. This is to be expected attending to the distribution of the confidences on the Figure 10 of the main document. }

\textcolor{black}{Since this metrics have been computed in a binary setting (cancer vs non-cancer) they relate to the binary accuracy presented in the main document.}

\begin{table}[h!]
\centering
\footnotesize	
\caption{Additional metrics measured on the Test set for the main methods evaluated in this document. We present the area under the precision recall (PR-AUC) curve, the area under the receiver operating characteristic (ROC-AUC) curve, and the F1-score.}
\label{tab:test_other_metrics}
\begin{tabular}{lccc}
\toprule
\textbf{Method}  & \textbf{PR-AUC} & \textbf{F1-Score} & \textbf{ROC-AUC}  \\ \midrule
iMIL4Path &  $ 0.996$  & $ 0.969$  & $0.984$\\
\textbf{Ours (CRS4K)} &  $0.994$  &$0.960$ & $0.978$  \\
\textbf{Ours (CRS10K) wo/ Agg} & $\textbf{0.998}$ & $\textbf{0.977}$ & $\textbf{0.992}$ \\
\bottomrule
\end{tabular}
\end{table}

\begin{table}[h!]
\centering
\footnotesize	
\caption{Additional metrics measured on the Prototype set for the main methods evaluated in this document. We present the area under the precision recall (PR-AUC) curve, the area under the receiver operating characteristic (ROC-AUC) curve, and the F1-score. The best values per column are in bold.}
\label{tab:prototype_other_metrics}
\begin{tabular}{lccc}
\toprule
Method  & PR-AUC & F1-Score & ROC-AUC  \\ \midrule
iMIL4Path &  $0.993$  & $0.959 $ & $0.979$ \\
Ours (CRS4K) &  $0.966$  &$0.926$  & $0.926$ \\
Ours (CRS10K) wo/ Agg & $\textbf{0.999}$ & $\textbf{0.980}$ & $\textbf{0.996}$ \\
\bottomrule
\end{tabular}
\end{table}

\begin{table}[h!]
\centering
\footnotesize	
\caption{Additional metrics measured on joint set of all the test datasets for the main methods evaluated in this document. We present the area under the precision recall (PR-AUC) curve, the area under the receiver operating characteristic (ROC-AUC) curve, and the F1-score. The best values per column are in bold.}
\label{tab:all_other_metrics}
\begin{tabular}{lccc}
\toprule
Method  & PR-AUC & F1-Score & ROC-AUC \\ \midrule
iMIL4Path &  $0.993$  & $0.943 $ & $0.965$ \\
Ours (CRS4K) &  $0.993$  &$0.965$ & $0.968$  \\
Ours (CRS10K) wo/ Agg & $\textbf{0.998}$ & $\textbf{0.982}$ & $\textbf{0.991}$ \\
\bottomrule
\end{tabular}
\end{table}

\begin{figure}[h!]
    \centering
    \includegraphics[width=\textwidth]{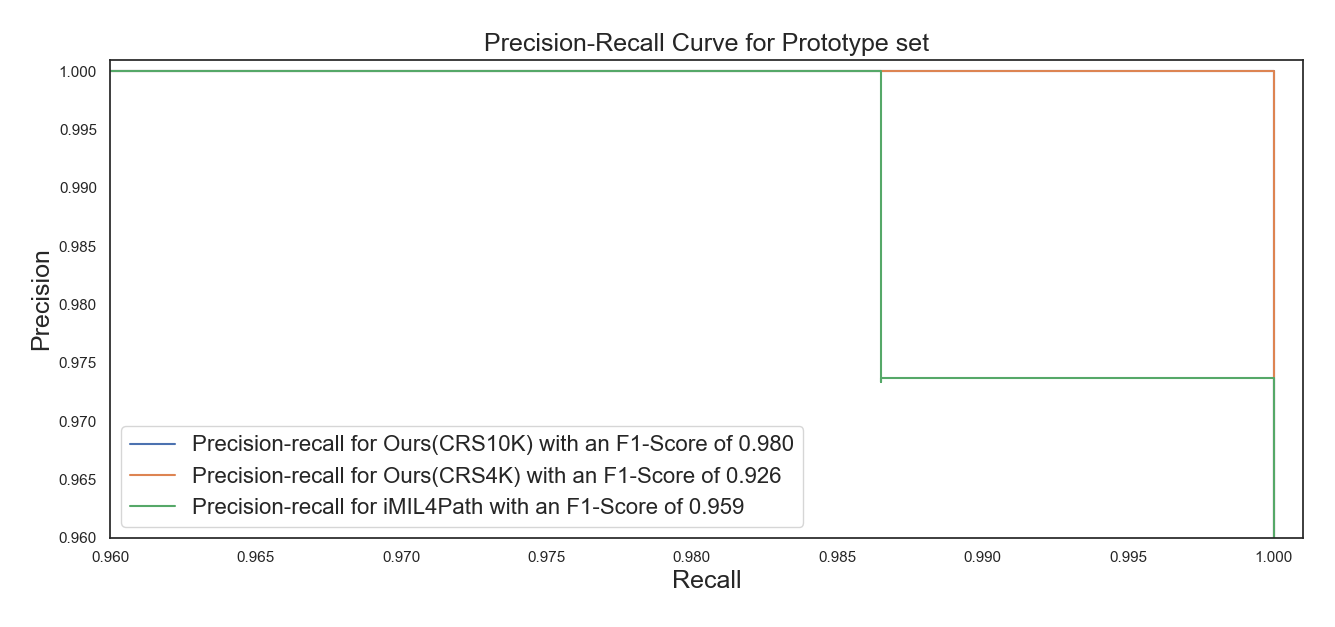}
    \caption{\textbf{Precision-recall curve on the on the Prototype set:} For the three distinct models, we have calculated the Precision-recall curve on this dataset. Includes an indication of the F1-Score for each of the different models.  The blue line represents the curve of Our method when trained on CRS10K, while the orange line shows the same method when trained on CRS4K. The green line is the curve of iMIL4Path.}
    \label{fig:proto_prc}
\end{figure}

Supplementary Figure~\ref{fig:proto_prc}\ shows the precision-recall curves for the three models evaluated on the prototype test set. Once again the F1-Score also highlights the performance of the proposed model to mitigate both errors.

\begin{figure}[h!]
    \centering
    \includegraphics[width=\textwidth]{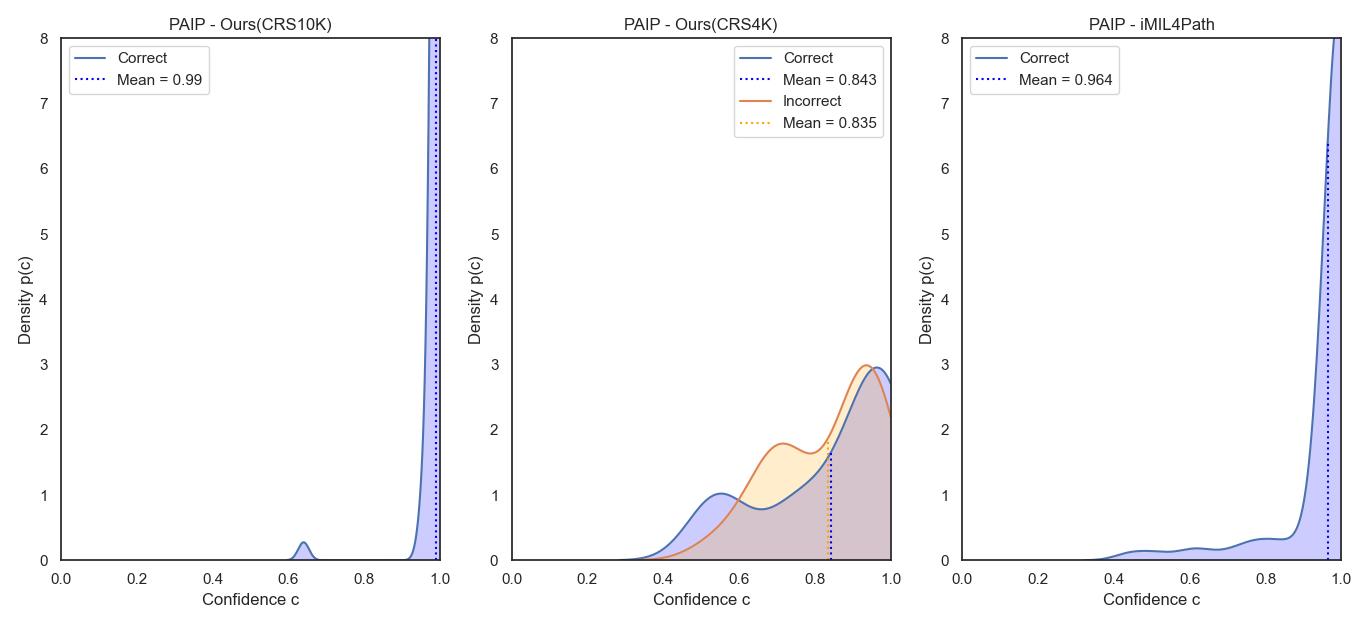}
    \caption{\textbf{Confidence analysis for correct and incorrect predictions on the PAIP dataset:} Kernel density estimation of the confidences of correct and incorrect predictions performed on the three-class classification problem by three distinct models on the PAIP dataset. The plots represent, from left to right, the proposed method trained on CRS10K, the proposed method trained on CRS4K and iMIL4Path. In each plot, the blue line defines the density function of the correct samples and the blue dashed line the mean confidence of those samples. On the other hand, the orange solid and dashed lines represent the same for incorrect predictions.}
    \label{fig:paip_confidences}
\end{figure}

Evaluation of the confidence of the model in the PAIP dataset shows that in two of the three approaches, the number of incorrect samples is one or zero, as such, there is no density estimation for wrong samples in their confidence plot as seen in Supplementary Figure~\ref{fig:paip_confidences}. Yet, it is visible the shift towards higher values of confidence in the proposed approach trained on the CRS10K when compared to the method of iMIL4Path. The version trained on CRS4K shows very little separability between the confidence of correct and incorrect predictions.

\begin{figure}[h!]
    \centering
    \includegraphics[width=\textwidth]{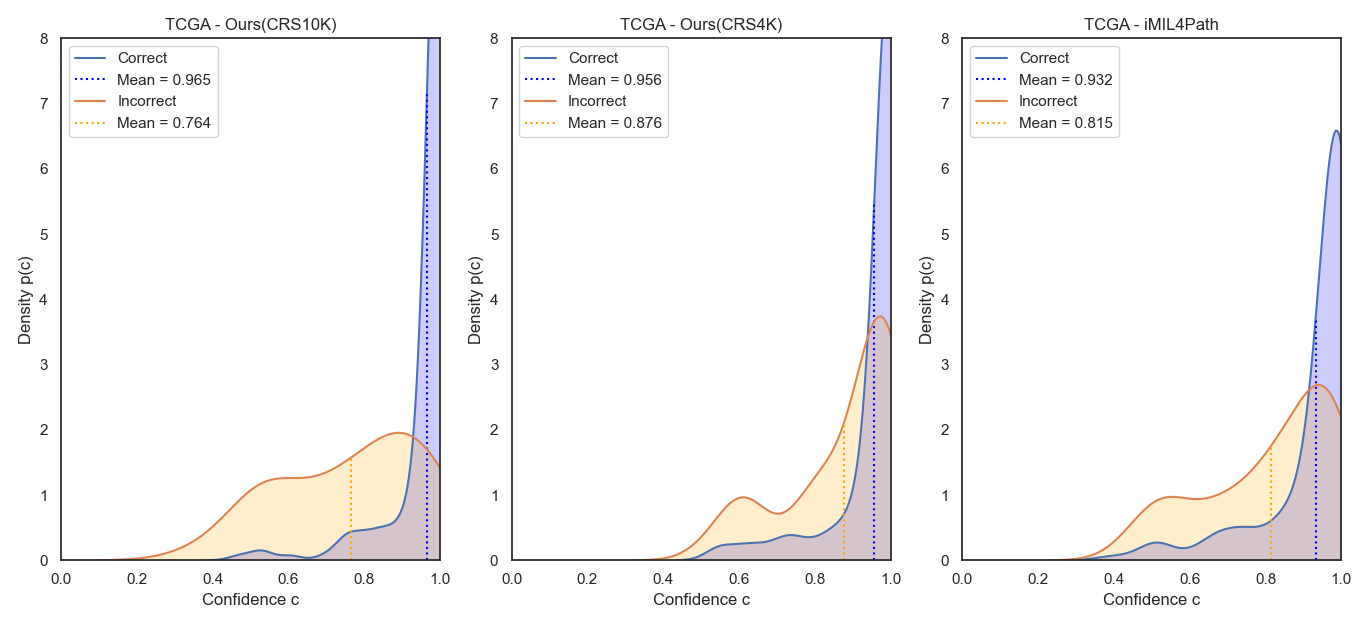}
    \caption{\textbf{Confidence analysis for correct and incorrect predictions on the TCGA dataset:} Kernel density estimation of the confidences of correct and incorrect predictions performed on the three-class classification problem by three distinct models on the TCGA dataset. The plots represent, from left to right, the proposed method trained on CRS10K, the proposed method trained on CRS4K and iMIL4Path. In each plot, the blue line defines the density function of the correct samples and the blue dashed line the mean confidence of those samples. On the other hand, the orange solid and dashed lines represent the same for incorrect predictions.}
    \label{fig:tcga_confidences}
\end{figure}

Inspecting the predictions' confidence for the three models, for the TCGA dataset, indicates a behaviour in line with the accuracy-based performance (Supplementary Figure~\ref{fig:tcga_confidences}). Moreover, a confidence shift of wrong predictions' confidence towards smaller values is clearly visible in the plot corresponding to the model trained on CRS10K.  The shown gap of 0.2 between the confidence of correct and wrong predictions, indicates that it is possible to quantify the uncertainty of the model and avoid the majority of the wrong predictions. In other words, when the uncertainty is above a learnt threshold, then the model refuses to make any prediction. It is extremely useful in models designed as a second opinion system.

\begin{figure}[h!]
    \centering
    \includegraphics[width=\textwidth]{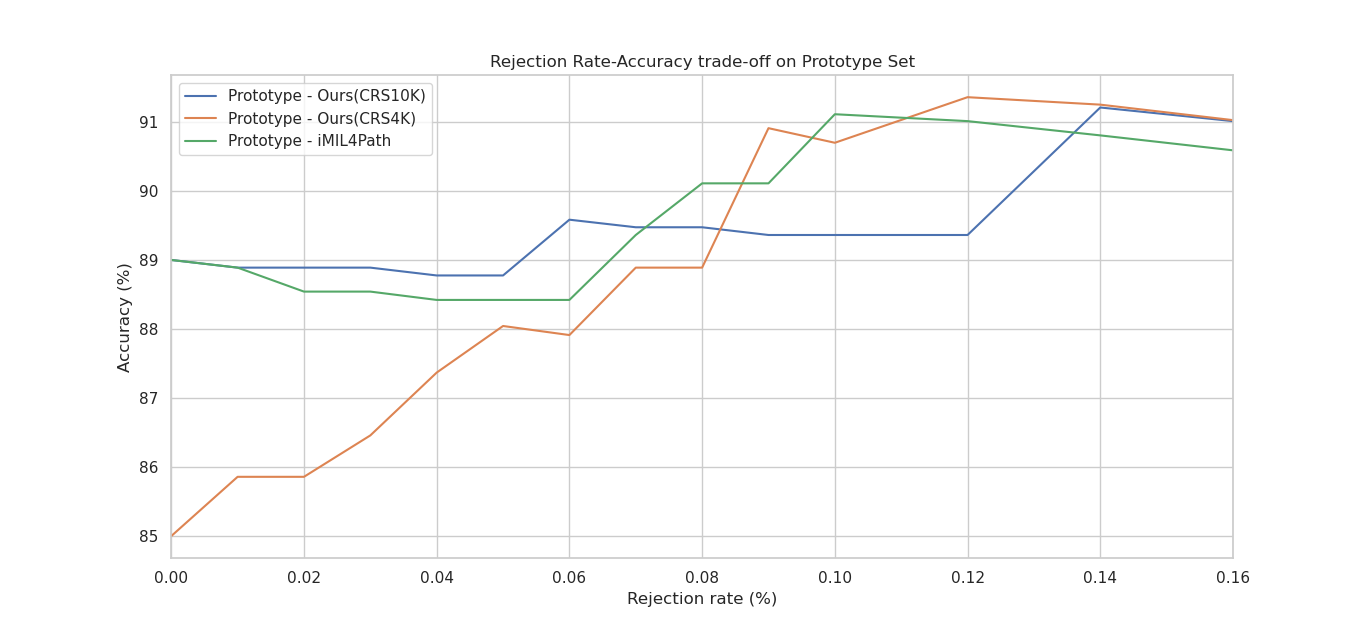}
    \caption{\textbf{Accuracy-vs-Rejection-rate for the models evaluated on the Prototype set}. Relation between the accuracy and the percentage of samples not classified by the model. Both axes are in percentage.  The blue line represents Our method when trained on CRS10K, while the orange line shows the same method when trained on CRS4K. The green line is for iMIL4Path.  }
    \label{fig:proto_rejection}
\end{figure}

On the prototype set (Supplementary Figure~\ref{fig:proto_rejection}), the gains are less evident, which can be assumed to be related to the lower data quantity. However, performances above 91\% are achieved by all the three models, with our model trained on the CRS10K data being the one that requires a larger rejection rate to achieve such value. Nonetheless, as the rejection rate increases the performance of that model improves to 100\% accuracy at less than 50\% rejection.

\begin{figure}[h!]
    \centering
    \includegraphics[width=\textwidth]{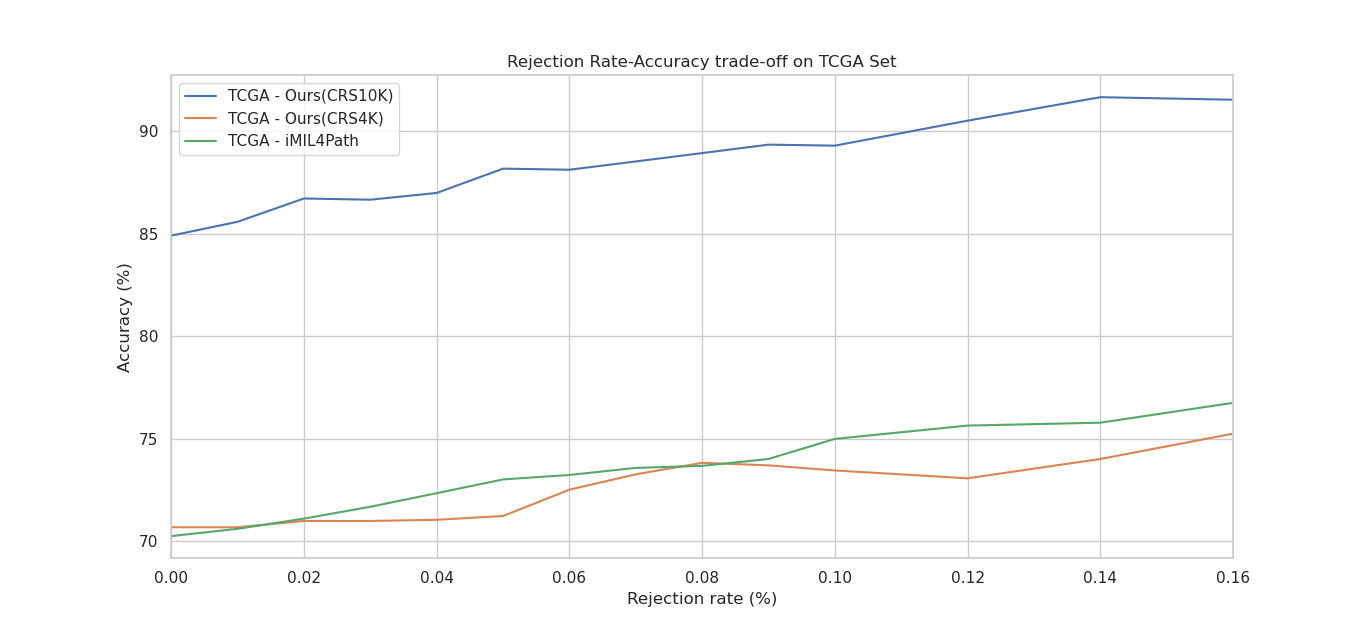}
    \caption{\textbf{Accuracy-vs-Rejection-rate for the models evaluated on the TCGA dataset}. Relation between the accuracy and the percentage of samples not classified by the model. Both axes are in percentage.   The blue line represents Our method when trained on CRS10K, while the orange line shows the same method when trained on CRS4K. The green line is for iMIL4Path}
    \label{fig:tcga_rejection}
\end{figure}

On the TCGA set, the performance is already significantly different without any rejection. The difference between the performance of the best performing model and the remaining is kept at the different rates. At 16\% rejection rate, as seen in Supplementary Figure~\ref{fig:tcga_rejection}, the accuracy of the model trained on the CRS10K dataset is 91.54\%. At 50\% the performance becomes 97.26\%.